\documentclass{article}

\font\scap=cmcsc10 \hfuzz=5cm
\font\scap=cmcsc10
\font\Bbb=msbm10

\usepackage{graphicx}

\newcount\eqnumber
\def\neweq{{\rm{(\the\eqnumber)}}\global\advance\eqnumber by 1}
\def\eqdef#1{\eqno\xdef#1{\the\eqnumber}\neweq}
\def\newaeq{{\rm{(\the\eqnumber {\it a})}}\global\advance\eqnumber by 1}
\def\eqdaf#1{\eqno\xdef#1{\the\eqnumber}\newaeq}
\def\eqdisp#1{\xdef#1{\the\eqnumber}\neweq}
\def\eqdasp#1{\xdef#1{\the\eqnumber}\newaeq}

\def\newaleq{{\rm{\the\eqnumber}}\global\advance\eqnumber by 1}
\def\eqdal#1{\xdef#1{\the\eqnumber}\newaleq}

\newcount\refnumber
\def\newref{{\the\refnumber}\global\advance\refnumber by 1}
\def\refdef#1{{\xdef#1{\the\refnumber}}\newref}

\begin{document}

\centerline{\bf Revisiting the Human and Nature Dynamics model}

\bigskip
\bigskip{\scap B. Grammaticos}
\quad{\sl IMNC,  CNRS, Universit\'e Paris-Diderot, Universit\'e Paris-Sud, Universit\'e Paris-Saclay, 91405 Orsay, France}

\medskip {\scap R. Willox} \quad
{\sl Graduate School of Mathematical Sciences, the University of Tokyo, 3-8-1 Komaba, Meguro-ku, 153-8914 Tokyo, Japan }

\medskip{\scap J. Satsuma}
\quad{\sl  Department of Mathematical Engineering, Musashino University, 3-3-3 Ariake, Koto-ku, 135-8181 Tokyo, Japan}

\bigskip
{\sl Abstract}
\smallskip

We present a simple model for describing the dynamics of the interaction between a homogeneous population or society, and the natural resources and reserves that the society needs for its survival. The model is formulated in terms of ordinary differential equations, which are subsequently discretised, the discrete system providing a natural integrator for the continuous one. An ultradiscrete, generalised cellular automaton-like, model is also derived. The dynamics of our simple, three-component, model are particularly rich exhibiting either a route to a steady state or an oscillating, limit cycle-type regime or to a collapse. While these dynamical behaviours depend strongly on the choice of the details of the model, the important conclusion is that a collapse or near collapse, leading to the disappearance of the population or to a complete transfiguration of its societal model, is indeed possible.



\bigskip
1. {\scap Introduction}
\medskip

Predicting the future has always been a major preoccupation of the human species. From shamanic rituals to parlour s\'eances, a huge amount of energy (and money) has been invested in, mostly futile, efforts at divining or mastering the future. The only promising answer to this conundrum is of course furnished by a scientific approach, using the tools of mathematical modelling, but it took until the advent of the modern computer before this approach became a viable one. 

One of the very first approaches to predicting the future evolution of human society, based on modelling, was that of Forrester [\refdef\forrest] in the framework of what he called ``system dynamics''. Already in the 1960s he developed models aimed at simulating the world's entire economic system. The collaboration of Forrester and his co-worker Meadows with the Club of Rome led to the publication of the --in some circles celebrated, in others infamous-- ``Limits to Growth'' [\refdef\meadows] monograph. The conclusions of Forrester and Meadows, in a nutshell, were that ``the world's economy tends to stop its growth and collapse as the result of a combination of reduced resource availability, overpopulation, and pollution''. This pessimistic view of the world's evolution is not new. It resonates with Malthus' conclusions [\refdef\malthus] in his ``Essay on the principle of population'' or, in a more recent past, with Hardin's in ``The tragedy of commons'' [\refdef\hardin]. As expected the Forrester-Meadows approach drew heavy criticism, and the results of their study  do not enjoy a wide acceptance despite the fact that recent studies, in particular those of Turner [\refdef\turner], do corroborate their predictions.

A major drawback of the Forrester World models is their extreme complexity. A model containing hundreds of dependent variables and thousands of parameters can easily be criticised despite the authors' claims on the robustness of their results.  It would therefore be interesting to have simple, manageable, models for which everybody can understand the basic assumptions, instead of intricate black-box-like ones, despite the latter's claim at enhanced realism. A first step in this direction was taken by Brander and Taylor [\refdef\brander] who proposed a model that involves just two interacting variables, the renewable resources and the population, for the description of the economics of Easter Island. The main result of these authors was a predator-prey-like behaviour of the system, with a periodically appearing pattern of ``feast and famine''. 

Recently, a paper entitled ``Human and Nature Dynamics'' (HANDY) [\refdef\handy] by Motesharrei, Rivas and Kalnay addressed the question of whether a complete societal collapse is, in fact, possible. Their approach is based on the modelling of structural inequalities in society, and which they implement by introducing two classes in the population: the `elites' and the `commoners'. This assumption was understood by many as a tacit political position and the study was seriously criticised, to the point that the article had to include a disclaimer by NASA making clear that it was ``an independent study by the university researchers utilising research tools developed for a separate NASA activity''. Still, the question of the possibility of a collapse of society is a vital one (all the more so, since the ``Limits to Growth'' studies point in the same direction) and one that cannot be easily  ignored.

In this study we will introduce a simple model situated halfway between the Brander-Taylor and the HANDY models. Our approach is a purely dynamical systems one, without reference whatsoever to political or societal questions. We shall study the conditions under which a system can undergo collapse or, instead, reach a steady (sustainable) state. While our initial model formulation is in terms of differential equations, we shall immediately proceed to a discretisation of the model, drawing upon the tools we have honed over the years [\refdef\mickens], and use the discrete equations as an integrator for the differential ones. One important question when considering discretisations is whether the scheme is a well-chosen one and not a scheme that only makes sense for very small discretisation steps. It is our belief that a sensible discretisation scheme must work for any discretisation step and even when the step goes to infinity. This limit is known as the ultradiscretisation procedure [\refdef\ud] and the resulting system is a generalised cellular automaton. The dynamics of the latter are considerably simplified compared to those of the initial discrete system, retaining only its essential features [\refdef\stefan]. Our study of the model will thus be completed by the introduction and simulation of its ultradiscrete counterpart.

\bigskip
2. {\scap Constructing the model}
\medskip

Our model comprises three interdependent variables, which for simplicity we shall refer to as `population', `resources' and `reserves'. The term population does not need any further explanation, but we insist on the presence of a single, unique, societal class that participates equally in production and consumption. The `resources' refer to renewable resources, which can be replenished by Nature and from which the population can extract means of subsistence and create growth through work. The term `reserves' is used for lack of a better one (`wealth' being even less appropriate) and it refers to existing non-renewable resources to which are added whatever resources that are produced from the renewable ones. 

The equations describing the dynamics of the system are written in terms of three variables, $x,y$  and $z$, which depend on $t$ (time) and which correspond  to the population, resources and reserves, respectively. These are, a priori, dimensional variables but we shall use all possible scalings of $x,y, z$ and $t$ to write the system, now effectively in terms of dimensionless variables, in the form:
$${dx \over dt}=A(r)x\eqdaf\zena$$
$${dy \over dt}=y(1-y)-xy\eqno(\zena b)$$
$${dz \over dt}=xy-B(r)x.\eqno(\zena c)$$
Here, $r$ stands for the intensive variable $r=z/x$. We believe that the use of an intensive variable, rather than the extensive variable $z$, is more appropriate in this type of model.

Equation (\zena a) describes the evolution of the population. We make the assumption that the rate of growth depends on the existing per capita reserves, hence the dependence on $r=z/x$. (Brander and Taylor [\brander] assume that this rate depends of the renewable resources but a dependence on the available reserves appears more reasonable).  We expect the general form of the function $A(z)$ to be like that given in Figure 1. In particular, we require that: $A(0)<0$ and $A'(r)>0$ for all $r\geq 0$.  In the following, we shall denote the unique zero of the function $A(r)$ by $\rho_0$, i.e. $A(\rho_0)=0$.

\centerline{\includegraphics[width=6cm,keepaspectratio]{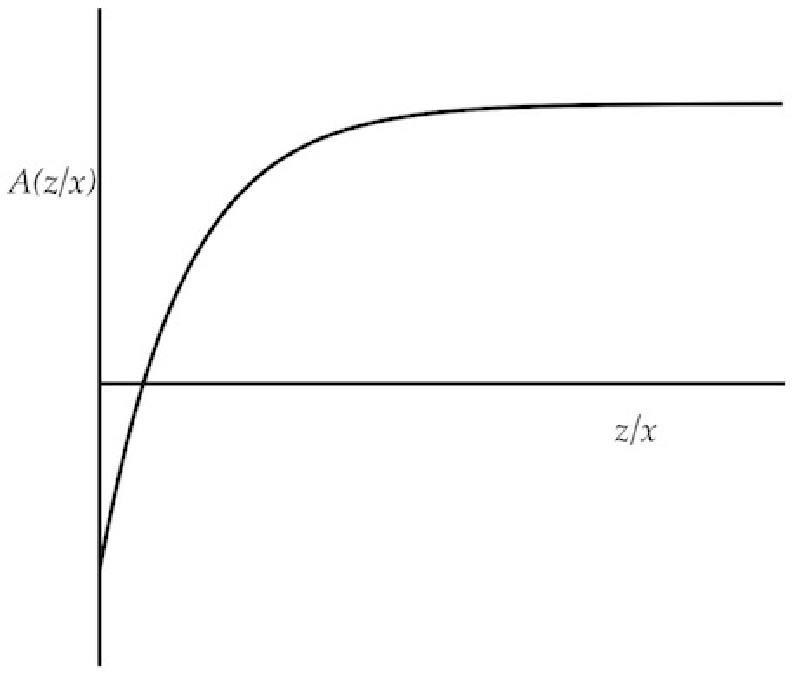}}
\vskip-.5cm\qquad\centerline{{Figure 1   }}\medskip

Clearly, when reserves become scarce the population dies out. On the contrary, an abundance of reserves leads to an increase in the birth-minus-death rate but, in practice, only up to a certain ceiling. As we shall see however, although it does make the model more realistic, this last assumption is not crucial to our conclusions. 

Equation (\zena b) describes the evolution of the renewable resources. In this equation we have used all possible scalings of $x,y$ and $t$ in order to put all coefficients to 1. The first term on the right-hand side of the equation describes the replenishing of resources by Nature, assuming a finite capacity of the world. The second term in the right-hand side corresponds to the extraction of resources by the work furnished by the population. A detailed justification of this equation can be found in the article of Brander and Taylor. 

Equation (\zena c) describes the accumulation and use of reserves. The first term in the right-hand side of this equation corresponds to the production of reserves out of renewable resources. (Its coefficient is put to 1 by an appropriate scaling of the variable $z$). Note that non-renewable resources can be included in the model by properly choosing the initial conditions $z_0$ for the variable $z$, which then effectively represent the amount of reserves that cannot be replenished. Finally, the second term in the right-hand side corresponds to the use of the reserves for subsistence and growth. 

\vskip.5cm
\centerline{\includegraphics[width=6cm,keepaspectratio]{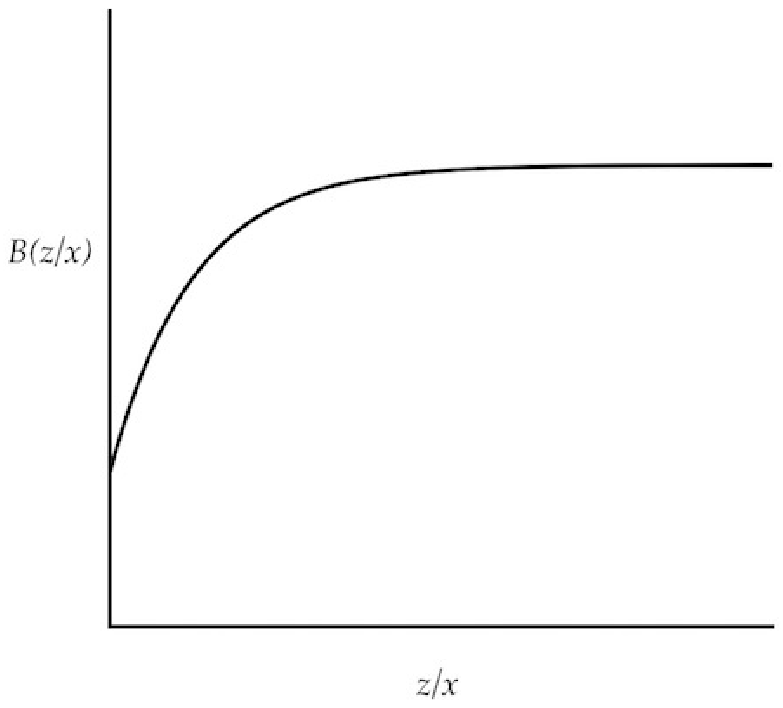}}
\vskip-.25cm\qquad\centerline{{Figure 2}}\vskip.25cm

The function $B(r)$ depends on the existing reserves and is of the form shown in Figure 2. 
Notice that $B(0)>0$ : the population needs a vital minimum for its bare subsistence. On the other hand, when the reserves become abundant one expects $B(z)$ to saturate lest the population sink into gluttony. Again, this last assumption, though realistic, is not central to our analysis. We do however require that $B'(r)>0$, for all $r\geq0$.

One important question is that of the positivity of the solutions to system (\zena), without which the model becomes irrelevant. Positivity is trivial for the $x$ and $y$ variables. If we start from an initial condition $x_0$ that is positive, the solution to (\zena a) will remain positive for all $t$ {as the derivative of $x$ vanishes when $x$ tends to 0.} Similarly, since $x$ is always positive, if we start from an initial condition $y_0\in \big]0,1\big[$, equation (\zena b) shows that $y$ will remain in this open interval at all times $t$. However, positivity for the $z$ variable is not automatically guaranteed. This problem is, in fact, best analysed on the evolution equation for the intensive variable $r=z/x$:
$${dr\over dt} = y - F(r),\eqdef\req$$
where
$$F(r) = B(r) + r A(r).\eqdef\Fdef$$
Clearly, $F(0)=B(0)>0$ and $F(\rho_0)=B(\rho_0)>0$. Moreover, we have $F'(r) >0$ whenever $r\geq\rho_0$,  but notice that, a priori,  $F'(0)= B'(0)+ A(0)$ can have either sign. In order to ensure positivity for the $r$ variable (and therefore for $z$) we shall -- besides the conditions on the functions $A(r)$ and $B(r)$ we already set out above -- impose one of the following two conditions, [C1] or [C2], depending on the precise situation we wish to study.

[C1] The functions $A(r)$ and $B(r)$ should be such that their combination $F(r)$, as given by (\Fdef), is a convex function of $r$ on an interval $\big[0,\rho_1\big]$ where $\rho_1$ is a zero of $F$: $F(\rho_1)=0$ (note that $\rho_1<\rho_0$, necessarily, since $F(r)>0$ for all $r\geq\rho_0$). For $r>\rho_1$ we only require $F'(r)>0$, allowing for a possible  inflection point beyond $\rho_1$.

In practice, condition [C1] tells us that $F'(0)<0$ and that $F(r)$ has in fact two zeros, $\rho_1$ and $\rho_2$ (counted with multiplicities: $F(\rho_1)=F(\rho_2)=0$ for $0<\rho_2\leq\rho_1<\rho_0$) and that it takes negative values in between $\rho_2$ and $\rho_1$, but  only positive ones beyond $\rho_1$. 
Then, for any value of $r$ in the interval $\big[\rho_2,\rho_1\big]$ where $F\leq0$, $y$ being positive, we see from equation (\req) that $r$ must increase and therefore always remains positive. Hence, under condition [C1], if we choose initial conditions $(x_0, y_0, z_0)$ such that $x_0>0, y_0>0$ and $z_0\gg x_0 \rho_2$, then  the solution to system (\zena) is guaranteed to be positive. Note that this effectively puts a lower bound on the amount of non-renewable resources that is needed to make the model meaningful. 

However, although its motivation looks quite intuitive, we will see in section 4 that the above condition is in fact too stringent for our purposes. Instead we may impose the following condition [C2], which generalises [C1] and which still ensures positivity for the $z$ variable though for more subtle reasons.

[C2] The functions $A(r)$ and $B(r)$ should be such that there is at least one positive value for $r$, say $\rho_*$, for which $F(\rho_*)=1$.

Clearly,  if $B(\rho_0)=F(\rho_0)\leq 1$ there necessarily exists such a $\rho_*$ in the interval $\big[\rho_0,+\infty\big[$. Moreover, since $A(r)<0$ when $r<\rho_0$, we have that in this case $F(r)<B(\rho_0) \leq1$ for all $r\in\big]0,\rho_0\big[$ and that, in fact, there is only one $\rho_*$ for which $F(\rho_*)$ takes the value 1 in the whole interval $]0,+\infty[$.
On the other hand, when $B(\rho_0)>1$, $\rho_*$ must lie in the interval $\big]0,\rho_0\big[$ and might not be unique. In that case, we define $\rho_*$ to be the greatest value of $r$ for which $F(r)=1$. In either case we have $F'(\rho_*)>0$ and starting from an initial condition $r_0\gg \max(\rho_0,\rho_*)$ we initially have $F(r_0)>1$ and $r$ will decrease, until it reaches a value $r<\rho_*$ for which $F(r)<y<1$, at which point the value of $r$ must increase again. Hence, if the variable $y$ does not approach 0 too closely but rather stays in the vicinity of 1, the positivity of $r$ (and $z$) is guaranteed even under the more relaxed condition [C2].

To see why the above scenario is indeed realised in this model we first need to perform a linear stability analysis, 
in full generality. Again, this analysis is best carried out on the system of equations (\zena a), (\zena b) and (\req), rather than (\zena c). First, depending on the precise condition we impose, [C2] or [C1], the model might have one or two fixed points that correspond to a `collapse', i.e. for which the value of $x$ is zero. 

If we take $x_*=0$ and look for a stationary solution to (\zena a), (\zena b) and (\req), then clearly $y_*=0$ or $y_*=1$ and $r_*$ must satisfy $F(r_*)=y_*$. Under condition [C2] there might not be any solutions to the equation $F(r_*)=y_*=0$, but if we impose condition [C1] then such a fixed point is guaranteed to exist and we have in fact two possibilities for 
$r_*$: $\rho_1$ or $\rho_2$. Both choices however correspond to $z_*=0$, yielding the fixed point $(x_*,y_*,z_*)=(0,0,0)$. 

When $x_*=0$ and $y_*=1$, the assumptions on $F(r)$ in condition [C1] and our choice of initial condition $r_0$ tell us that $r\geq\rho_2$ and that there is exactly one solution to the constraint $F(r_*)=1$ which, moreover, must lie beyond $\rho_1$: $r_*>\rho_1$. 
On the other hand, if we impose condition [C2], the existence of a solution to $F(r_*)=1$ is also guaranteed but the exact position of the point $\rho_*$ will depend on the value of $B(\rho_0)$, as explained above. Either way, we see that conditions  [C1] and [C2] both guarantee the existence of  the fixed point  $(x_*,y_*,z_*)=(0,1,0)$.
 
 Linearising the equations (\zena a), (\zena b) and (\req) around a fixed point $(x_*, y_*, r_*)$, we find
 $${d\xi\over dt}=A(r_*) \xi + x_* A'(r_*)\zeta\eqdaf\zdyo$$
$${d\eta\over dt}=-y_*\xi+(1-2y_*-x_*)\eta\eqno(\zdyo b)$$
$${d\zeta\over dt}=\eta-F'(r_*)\zeta.\eqno(\zdyo c)$$

 For the collapse fixed points, for which $x_*=0$, the eigenvalues that determine the linear stability of the fixed point can be simply read off from this system of linear equations: they are $A(r_*), 1-2y_*$ and $-F'(r_*)$ {(the first and last one of which govern the stability in the $x$ and $r$ directions respectively).} When a fixed point $(x_*,y_*,r_*)=(0,0,\rho_j)$ ($j=1,2$) exists (in the case of condition [C1] or, in certain cases, under condition [C2]) these eigenvalues become $A(\rho_j)$, $1$ and $-F'(\rho_j)$, the first one of which is always negative as $\rho_j<\rho_0$ but the last one of which is positive for $\rho_2$ and negative for $\rho_1$. However, reconstituting the linear perturbation for $z=r x$, we see that the behaviour of $z$ around $0$ is $z\approx 0 + \rho_j \xi_0 e^{A(\rho_j)} + \xi_0\zeta_0 e^{A(\rho_j) -F'(\rho_j)}$, i.e. always attractive as not only $A(\rho_j)<0$ but also $A(\rho_j) -F'(\rho_j) = -(B'(\rho_j) + \rho_j A'(\rho_j))<0$ for both $j=1,2$. {The fixed point $(x_*,y_*,z_*)=(0,0,0)$, though repulsive in one direction,  is therefore always attractive in the $x$ and $z$ directions.}
 
For the collapse fixed point with $y_*=1$ we find eigenvalues $A(r_*), -1$ and $-F'(r_*)$. Now, as we saw, under condition [C2] we have $r_*=\rho_*$ and $-F'(\rho_*)<0$. Likewise, under condition [C1] we have $r_*>\rho_1$ and since $F(r)$ increases monotonically for $r>\rho_1$ we know that $-F'(r_*)<0$ as well. 
Moreover, as explained above, we find that if $F(\rho_0)=B(\rho_0)<1$ then $r_*$ must lie beyond $\rho_0$ and that, on the contrary, if $B(\rho_0)>1$ we have $r_*<\rho_0$ (and $\rho_1<r_*<\rho_0$ under condition [C1]). Hence, if $B(\rho_0)>1$, the fixed point $(x_*,y_*,z_*)=(0,1,0)$ is attractive in all three directions but otherwise it is repulsive in exactly one direction (with the exception of the case $B(\rho_0)=1$ for which it is neutral in that direction). 
 
Most importantly however, when the above fixed point is unstable there actually exists a third fixed point, corresponding to a {\sl steady} state for the system. Indeed, when $B(\rho_0)<1$, if we take $r_*=\rho_0$ we see that $y_*=B(\rho_0)$ and $x_*=1-B(\rho_0)$ not only provide a stationary solution to the system (\zena a), (\zena b), (\req), but also fall into the physical range we impose for these variables: $x_*>0$ and $y_*\in\big]0,1\big[$. This situation corresponds to the fixed point $(x_*,y_*,z_*)=(1-B(\rho_0),B(\rho_0),(1-B(\rho_0)) \rho_0)$ for (\zena a)$\sim$(\zena c). 
The linear stability analysis for this fixed point is slightly more complicated than for the previous two cases. From the linear system (\zdyo a)$\sim$(\zdyo c), we obtain the characteristic equation
$$\lambda^3+(B(\rho_0)+F'(\rho_0)) \lambda^2+ B(\rho_0) F'(\rho_0) \lambda+ B(\rho_0) A'(\rho_0) (1-B(\rho_0))=0,\eqdef\ztri$$
for the eigenvalues that determine the linear stability for $(x_*,y_*,r_*)=(1-B(\rho_0),B(\rho_0),\rho_0)$.
Since all coefficients in the polynomial in (\ztri) are positive we find that it has no positive roots but at least one negative one. If all three roots are negative, or if the real part of the two (conjugate) complex roots is negative, we find that the `steady state' fixed point is attractive. The Routh-Hurwitz criterion tells us that this is the case if (and only if)
$$F'(\rho_0)(B(\rho_0)+F'(\rho_0)) > A'(\rho_0) (1-B(\rho_0)).\eqdef\negcond$$
However, if the inequality in this condition goes the other way, the real part of the complex roots is positive and the fixed point is unstable, which leaves open the possibility that there might be a limit cycle for the system. 

This brings us  back to the question of positivity under the condition [C2]. As we saw, if $B(\rho_0)>1$ the fixed point $(x_*,y_*,z_*)=(0,1,0)$ is attractive and positivity is almost surely guaranteed. If however, $B(\rho_0)<1$ the picture becomes more complicated and, in the following, we shall rely on the results of numerical simulations to verify the positivity of the solutions for  some special realisations of the model. These simulations will be performed with the help of integrators we shall construct by ad hoc discretisations of the model. Note that although the intensive variable $r=z/x$ proved to be very useful in the above general analysis, it is clear that $z$ is a more natural  variable and we shall therefore, hereafter, present all simulation results for the triplet of variables $(x,y,z)$.

\goodbreak
3. {\scap Special realisations of the model}
\medskip

The first realisation we are going to consider is that of a natalist, conservative, society. Namely we shall take the function $A(r)$ to be linear of the form $A(r)=\alpha r-\beta$, for positive constants $\alpha$ and $\beta$, which means that the birth rate increases unchecked in pace with the availability of reserves. Moreover we shall choose $B(r)=\gamma$, a positive constant, meaning that every individual consumes always the same amount of reserves. Clearly this is a toy model without any aspiration to realism. Still, it is the perfect example on which to explain the discretisation procedure and the construction of the integrator we shall use. The equations of the model now become 
$${dx\over dt}=\alpha z-\beta x\eqdaf\ztes$$
$${dy\over dt}=y(1-y)-xy\eqno(\ztes b)$$
$${dz\over dt}=x(y-\gamma).\eqno(\ztes c)$$
For these $A(r)$ and $B(r)$ we have $F(r)=\alpha r^2 -\beta r + \gamma$, which is convex for any choice of parameters, but condition [C1] imposes that $F(r)$ must have a zero ($\rho_1$), which forces us to require the condition $\beta^2\geq 4 \alpha\gamma$ to be satisfied. We then find that the initial condition for $z$ should satisfy $z_0 > x_0 (\beta-\sqrt{\beta^2- 4 \alpha\gamma})/(2\alpha)$. 
If, instead, we choose to relax the above condition by imposing condition [C2] we have to require that $\beta^2>4\alpha(\gamma-1)$, which is trivially satisfied whenever $\gamma\leq1$. In this case we find that the initial condition for $z$ should satisfy $z_0 > x_0 (\beta+\sqrt{\beta^2- 4 \alpha(\gamma-1)})/(2\alpha)$.
Note also that we have $\rho_0=\beta/\alpha$.

On these equations it is easily checked that, as explained above, the fixed point $(0,0,0)$ (when it exists) is always repulsive in the $y$ direction (while the remaining two eigenvalues are always negative) and that the fixed point $(0,1,0)$ is stable only if $B(\rho_0)=\gamma>1$. When $\gamma<1$, the steady state fixed point $(1-\gamma, \gamma, \beta(1-\gamma)/\alpha)$ appears, which is stable if and only if $\gamma > (\alpha-\beta^2)/(\alpha+\beta)$.

In order to discretise the equations of system (\ztes) we shall apply a prescription that can be summarised as: ``if the dependent variables of an equation correspond to positive quantities, the discretisation should be such as to avoid any minus signs''. In the present case this can be achieved as follows.

Let us discretise the independent (time) variable $t$ as $t_n=t_0+nh$, where $h$ is a small (positive) time step, and let us introduce the notation $x_n\equiv x(t_n), y_n\equiv y(t_n)$ and $z_n\equiv z(t_n)$. For the discretisation of the time derivative we simply  take ${dx\over dt}\approx (x_{n+1}-x_n)/h$. Then, in order to avoid any minus signs in the resulting discrete evolution equations, we stagger the first two equations as:
$${x_{n+1}-x_n\over h}=\alpha z_n-\beta x_{n+1}\eqdaf\zpen$$
$${y_{n+1}-y_n\over h}=y_n-y_ny_{n+1}-x_ny_{n+1}.\eqno(\zpen b)$$
However, no staggering can bring equation (\ztes c) to a discrete form without minus signs. The way to deal with this difficulty is to multiply the $\gamma x$ term by a fictitious factor $z/z$ and then apply the adequate staggering, as in:
$${z_{n+1}-z_n\over h}=x_ny_{n+1}-\gamma x_n z_{n+1}/z_n.\eqno(\zpen c)$$
Note that the $xy$ term was discretised to $x_ny_{n+1}$ so as to coincide with the term present in (\zpen b). 

We then obtain the discrete system
$$x_{n+1}={x_n+h\alpha z_n\over1+h\beta}\eqdaf\zhex$$
$$y_{n+1}={(1+h)y_n\over1+h(y_n+x_n)}\eqno(\zhex b)$$
$$z_{n+1}=z_n{z_n+hx_ny_{n+1}\over z_n+h\gamma x_n}.\eqno(\zhex c)$$
We remark that, thanks to our appropriate staggering, no minus signs appear in the discrete equations and thus, starting from positive initial conditions and positive parameters, it is guaranteed that the iteration of (\zhex) only yields positive values at all times. Notice also that, in order to ensure positivity through the introduction of the $z_{n+1}/z_n$ term, we end up with equations that are not time-reversible. This is however a very small price to pay since our main interest will lie in the approach of this system to an equilibrium or a collapse.

Given the way in which the discretisation was implemented, it should be clear that the fixed points of the map defined by (\zhex a)$\sim$(\zhex c) are exactly those of the continuous system (\ztes a)$\sim$(\ztes c). However, since the positivity of the iterates of this map is now guaranteed by construction, whereas that of the solutions to the original continuous system required a specific condition on the parameters (either $\beta^2>4\alpha\gamma$ or $\beta^2>4\alpha(\gamma-1)$) to be satisfied, one could wonder whether this discretisation will be faithful to the behaviour of the original system for any choice of $\alpha, \beta$ and $\gamma$. Let us therefore analyse the stability of the fixed points for the map (\zhex).

Linearising (\zhex) around $(x_*,y_*,z_*)=(0,0,0)$, under condition [C1] i.e. $\beta^2>4\alpha\gamma$, we find the following characteristic equation for the eigenvalues for the linear perturbation:
$$\left(\lambda -(1+h)\right) \left(  (1+\beta h) \lambda^2 - (2+\beta h) \lambda + 1 + \alpha \gamma h^2\right) =0.\eqdef\chpolyuno$$
The eigenvalue $\lambda=1+h$ (which is that for the perturbation in the $y$-direction) is greater than 1 and this fixed point is therefore always repulsive in the $y$-direction, just as in the continuous system (\ztes). Since the remaining quadratic polynomial in (\chpolyuno) is always positive when $\lambda=\pm1$, it is easily verified that its roots will both have modulus less than 1 if and only if $(1+\alpha\gamma h^2)/(1+\beta h) <1$,  i.e. if $\beta>\alpha\gamma h$. For fixed values of $\alpha, \beta$ and $\gamma$ this yields a constraint on the discretisation step $h$ under which the fixed point $(0,0,0)$ is attractive in the remaining two directions, as was the case for the continuous system. Note that as this fixed point might still exist even if we choose to impose condition [C2] instead of [C1], we shall require the constraint $\beta>\alpha\gamma h$ to be satisfied, even if we work under condition [C2]. In fact, under this condition on the parameters the behaviour of the discretisation near its fixed points will faithfully mimic that of the original system. 

For $(x_*,y_*,z_*)=(0,1,0)$, we find the characteristic equation 
$$\left( (1+h) \lambda -1\right) \left(  (1+\beta h) \lambda^2 - (2+\beta h) \lambda + 1 + \alpha (\gamma-1) h^2\right) =0.\eqdef\chpolydue$$
The $y$-direction of the linear perturbation around this fixed point is attractive since $\lambda=1/(1+h)<1$. Furthermore, as the remaining quadratic polynomial in (\chpolydue) is negative at $\lambda=1$ when $\gamma<1$, we know that in this case it always has a root that is greater than 1 and that this fixed point will be unstable. Just as in the continuous case, $\gamma<1$ is exactly the condition on $\gamma$ for which the fixed point $(1-\gamma, \gamma, \beta(1-\gamma)/\alpha)$ appears. The linear stability analysis for this last  fixed point is too involved to be given here but suffice it to say that there is always at least one attractive root (i.e. with modulus less than 1) but that the remaining roots are either both attractive or both repulsive, just as in the continuous case. 

Taken together with the explanations in section 2 we can therefore say that, near the continuum limit, if we impose conditions [C1] or [C2] on the discrete system as well, the evolution in our discrete system  is in fact faithful to that of the continuous system, as long as the orbit of an initial condition stays in the vicinity of the point $(x,y,z)=(0,1,0)$ or, more generally, if it does not `hit' the $xy$-plane transversally at $(x,y,z)=(0,0,0)$. In section 4 it will be shown on a specific numerical example that this last condition can indeed be checked.

The toy model (\ztes) was introduced essentially in order to illustrate the construction of the discrete system which will be used as an integrator for the differential one. Clearly, the behaviour of the functions $A$ and $B$ we assumed for (\ztes) is not very realistic and a better assumption is therefore in order. To this end we shall use the results of Holling  [\refdef\holling] on functional response. Holling's theory is based on the assumption of a decelerating intake rate, which in our case would mean that an increase of reserves leads to a slowing-down of the increase in both birth rate and consumption. This effect is often modelled by what Holling called the disc equation, which has the form $f(s)=s/(\lambda+s)$ where $f(s)$ is the response to an increase of the resource $s$. In the present case we shall use the disc equation to model the response to an increase of $z/x$. In particular, we choose the functions $A(r)=(\alpha r-\beta)/(r+\lambda)$ and $B(r)=\gamma+\delta r /(\mu+r)$ in terms of three additional (positive) parameters $\lambda, \delta$ and $\mu$,
leading to the system
$${dx\over dt}={\alpha z-\beta x\over \lambda x+z}x\eqdaf\zhep$$
$${dy\over dt}=y(1-y)-xy\eqno(\zhep b)$$
$${dz\over dt}=xy-\gamma x-{\delta zx\over \mu x+z}.\eqno(\zhep c)$$
There is no need to repeat the fixed point and stability calculations for system (\zhep) since we can always find parameter values for which the functions $A(r)$ and $B(r)$ conform to the conditions [C1] or [C2] of section 2 and the properties of the system can therefore be deduced from the general results in that section. However, because of the additional parametric freedom in (\zhep) it is interesting to look at some special limits. Taking $\lambda\gg1$ and $\delta=0$ we obtain precisely (\ztes) (after a renormalisation of the remaining constants). On the other hand taking $\lambda\ll 1$ gives for (\zhep a) an equation where the death rate increases unbounded when the reserves become scant (but this is a rather unrealistic assumption). Finally, when we take $\mu\gg1$ and $\delta$ of the same order of magnitude, we obtain an insatiable consumption that keeps increasing along with the reserves, which is an equally unrealistic situation. 

The discretisation of (\zhep) is straightforward when one follows the same prescriptions we used for the derivation of (\zhex). We obtain
$$x_{n+1}=x_n{\lambda x_n+(1+h\alpha)z_n\over(\lambda+h\beta) x_n+z_n}\eqdaf\zoct$$
$$y_{n+1}={(1+h)y_n\over1+h (y_n+x_n)}\eqno(\zoct b)$$
$$z_{n+1}=z_n{z_n+hx_ny_{n+1}\over z_n+h\gamma x_n+h\delta z_nx_n/(\mu x_n+z_n)}.\eqno(\zoct c)$$
The fixed points of the discrete system are again exactly the same as for the continuous one. Finding a (manageable) condition on $h$ such that the stability properties of these fixed points are the same as for the continuous system, {in full generality,  might be too involved  to be practical but is quite straightforward for specific values of the parameters.}

The discretisation of the model (\zena) for more general functions can also be achieved in a similar fashion. The first step is always to separate the functions $A$ and $B$ into a positive and a negative part and then to apply the proper staggering in order to set up the discrete system, so as to avoid any minus signs appearing. Consider for example the functions represented in Figures 1 and 2, which were obtained from the ansatz $A(r)=\alpha-\beta\exp(-\lambda r)$ and $B(r)=\gamma-\delta\exp(-\mu r)$ where $\beta>\alpha$ and $\gamma>\delta$. The separation into positive and negative parts is obvious but note that, in general, in the case of (\zena c) a multiplication by a $z/z$ factor may be necessary. In this case, the system thus obtained takes the form:
$$x_{n+1}={(1+\alpha h) x_n\over 1+\beta h e^{-\lambda z_n/x_n}}\eqdaf\zenn$$
$$y_{n+1}={(1+h)y_n\over1+h (y_n+x_n)}\eqno(\zenn b)$$
$$z_{n+1}=z_n{z_n+hx_ny_{n+1}+h\delta x_n e^{-\mu z_n/x_n}\over z_n+h\gamma x_n}.\eqno(\zenn c).$$

In what follows we shall present simulations of our model based on the systems (\ztes), (\zhep) as well as for the parametrisation with exponentials.

\bigskip
4. {\scap Simulation results}
\medskip

The simulations we are going to present are based on the discrete systems (\zhex), (\zoct) and (\zenn). In all simulations the time step $h$ was taken to be 0.01, which gives a solution we expect to be very close to that of the differential system. We have checked however that modifying $h$ by a factor of 10 either way does not qualitatively alter the results.  In all simulations we took $x_0=0.01$, $y_0=1.0$ and $z_0=0.5$ as an initial condition. These values correspond to a population emerging from a very small nucleus, a pristine state for the natural resources and rather abundant (non-renewable) reserves, a choice which sounds reasonable given that we are aiming at describing the human race and its interaction with Nature. The value we took for $z_0$ is rather arbitrary but we do not expect it to qualitatively change our conclusions. Again, variations of the initial values also do not have an impact on the final state of the system.  

In all the figures in this section the evolution of the $x$ variable is plotted with a solid line, that of $y$ with a dashed line and that of $z$ with a dash-dot line.

We start with simulations of the very simplistic model (\ztes). The first figure shows a collapse situation. 

\vskip.5cm
\centerline{\resizebox{10cm}{!}{\includegraphics{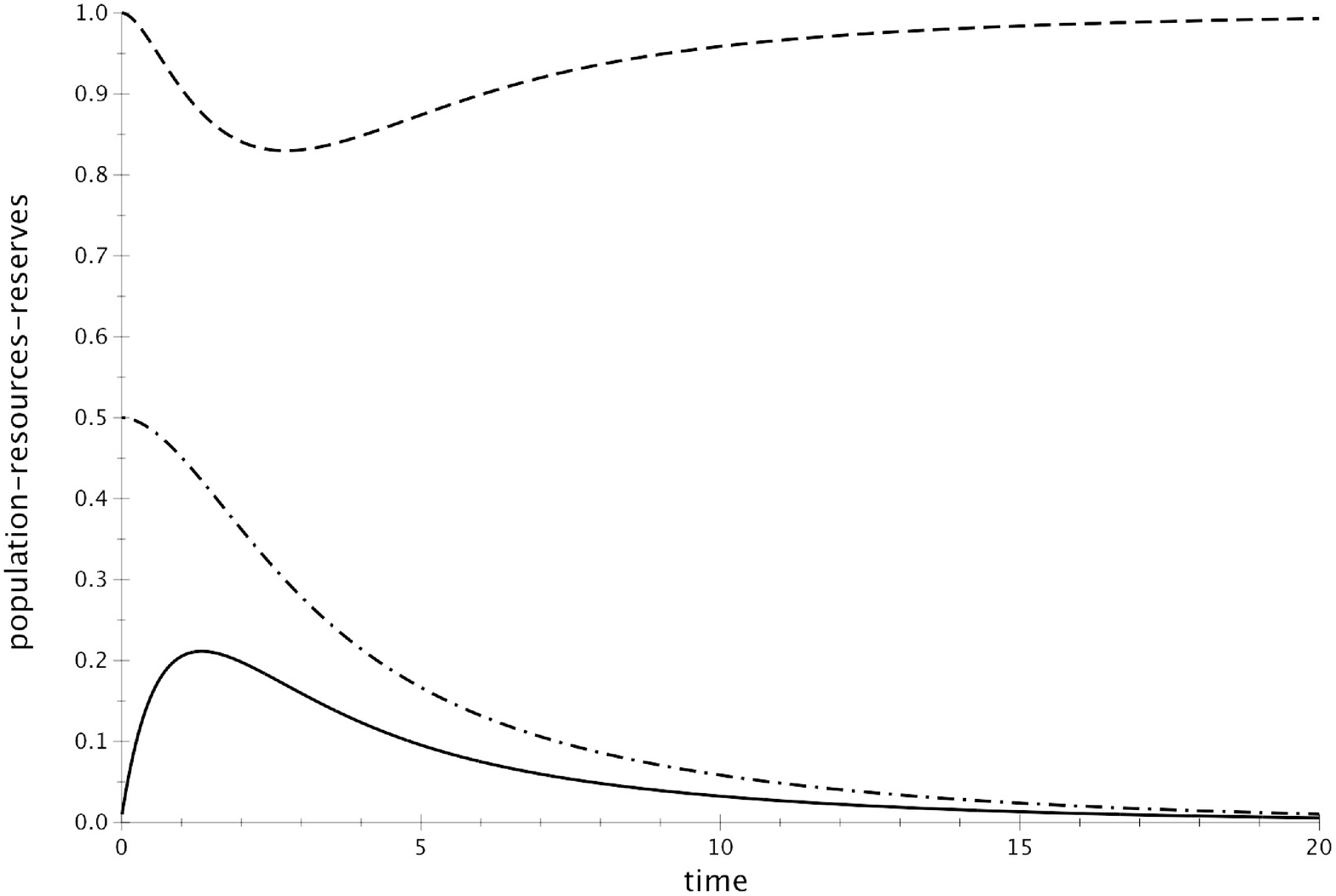}}}\vskip-.25cm\qquad\centerline{{Figure 3. Simulation results for system (\zhex) for $\alpha=1.0, \beta=2.0$ and $\gamma=1.3$.}}\vskip.5cm

A major flaw of the model is immediately visible: the population growth has a negative curvature, something quite unrealistic. Typically one would expect the initial growth phase to resemble an exponential. Moreover, for all choices of parameters that satisfy the positivity conditions ([C2] in this particular case) we obtain a collapse that corresponds to a situation of `prolonged agony' where the population goes to zero very slowly, which is something that, somehow, does not look very realistic. 

The only way to `hasten' the collapse in this model would be to deliberately choose parameter values that violate the positivity condition, as is shown in the next figure.

\vskip.5cm
\centerline{\resizebox{10cm}{!}{\includegraphics{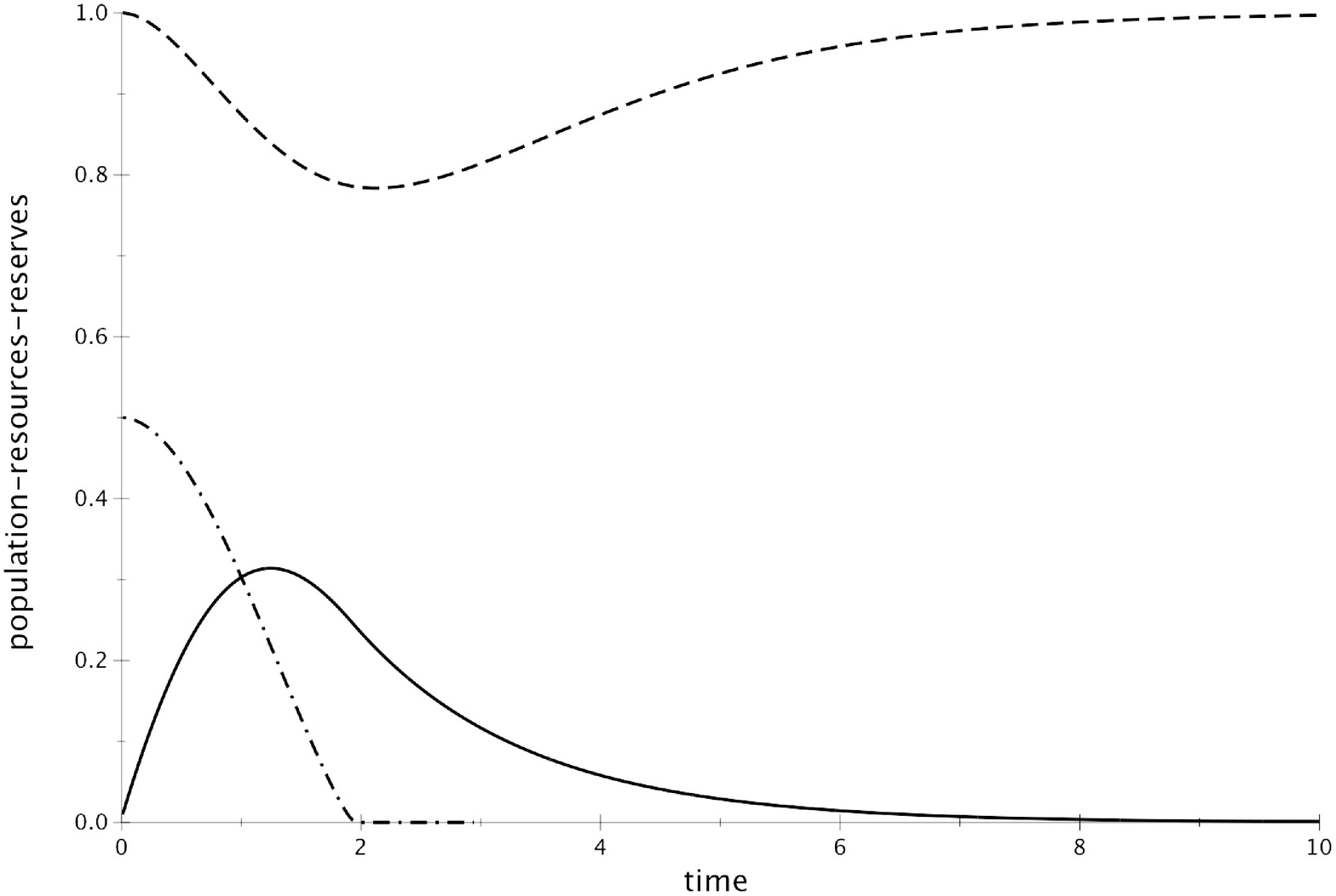}}}\vskip-.25cm\qquad\centerline{{Figure 4. Simulation results for system (\zhex) for $\alpha=1.0, \beta=0.70$ and $\gamma=2$.}}\vskip.5cm

Notice that the curve for the reserves $z$, when it approaches the horizontal axis, has the tendency to cross over to negative values and that it only levels off and falls to zero at the very last moment. This is of course due to the fact that the discretisation we have adopted enforces positivity, but had we not taken these discretisation precautions this choice of parameters would have led to negative values for $z$ and this particular collapse scenario should therefore be deemed unphysical.

However a collapse is not the only possible outcome. By changing the values of the parameters of the model as in Figure 5 (this time satisfying the positivity condition [C1]) it is possible to obtain a transition towards a steady-state. Note that it is possible to choose the  parameters such that the oscillations one observes here are washed out and the approach to the steady-state becomes monotonic.

\vskip.5cm
\centerline{\resizebox{10cm}{!}{\includegraphics{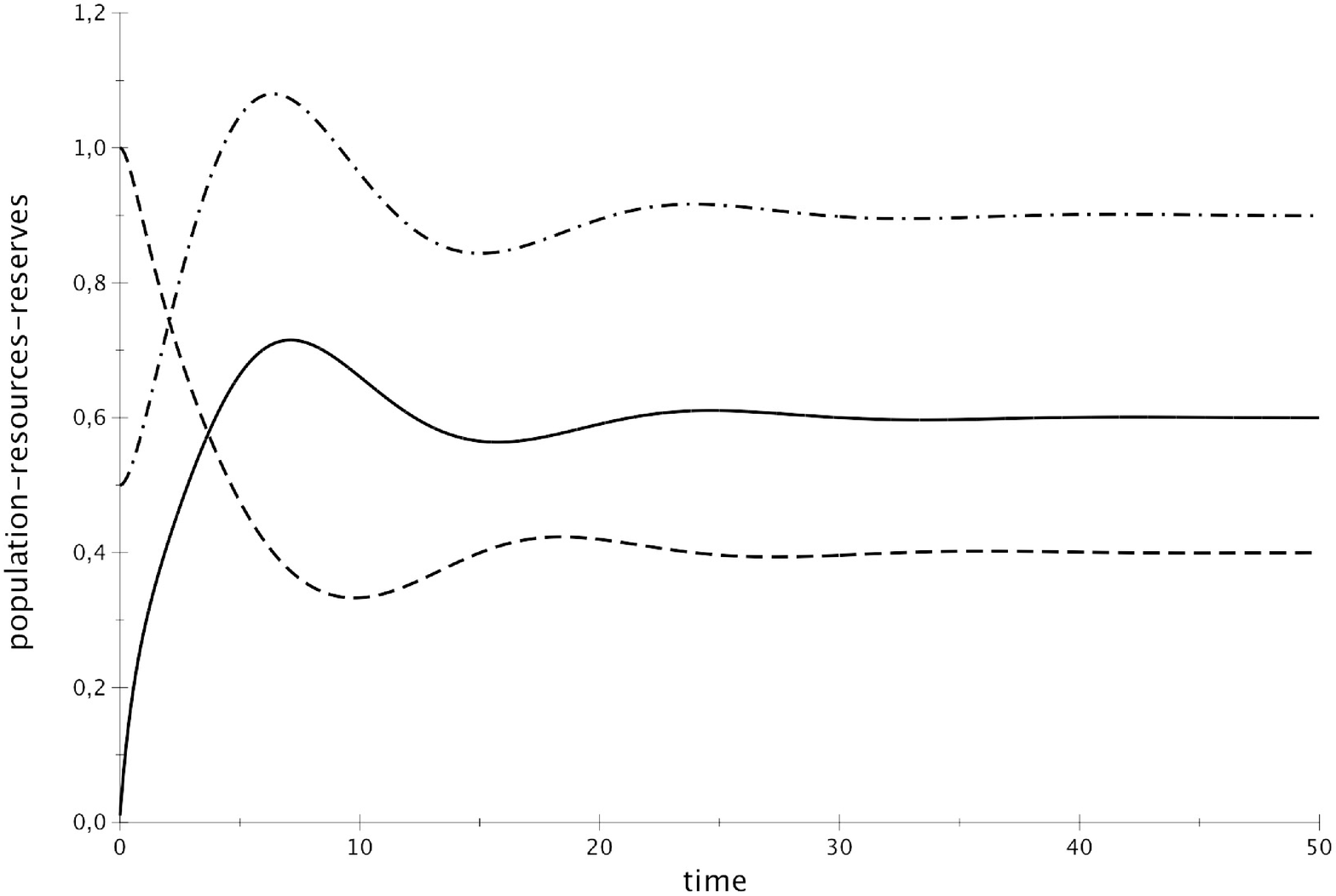}}}\vskip-.25cm\qquad\centerline{{Figure 5. Simulation results for system (\zhex) for  $\alpha=1.0, \beta=1.5$ and $\gamma=0.4$.}}\vskip.5cm

We move on to a more realistic model, system (\zhep), where both functions $A$ and $B$ saturate thanks to the parametrisation in terms of Holling's disc-equation.  Figure 6 shows a collapse situation. Notice that now the population growth has initially a positive curvature, as expected. 

\vskip.5cm
\centerline{\resizebox{10cm}{!}{\includegraphics{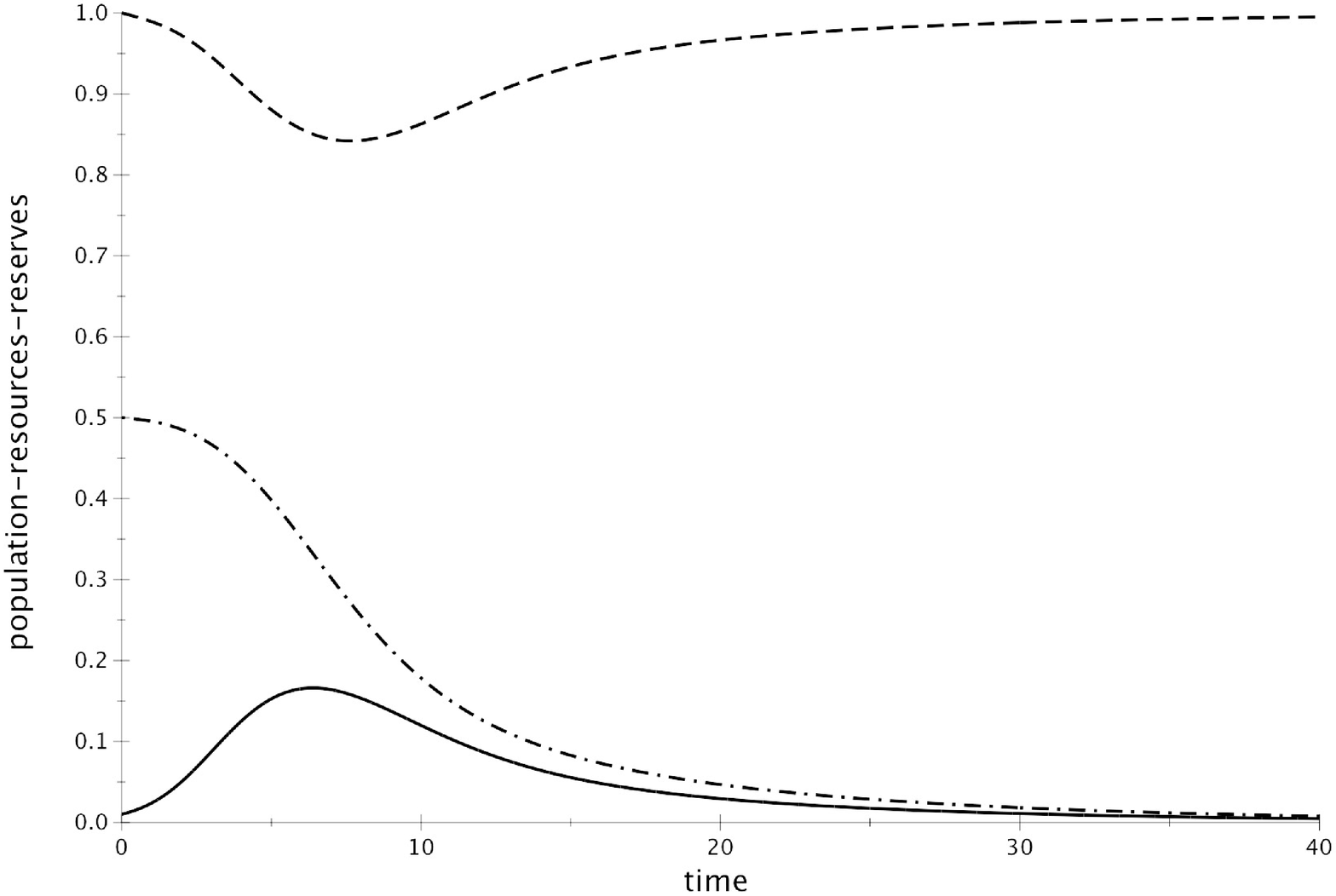}}}\vskip-.25cm\qquad\centerline{{Figure 6. Simulation results for system (\zoct) for $\alpha=1.0, \beta=2, \gamma=1, \delta=0.3, \lambda=2$ and $\mu=2$.}}\vskip.5cm

However the `slow death' scenario is still present, something we do not seem to be able to avoid if we enforce the positivity constraints on the parameters. A non-collapse steady state is also possible within this model, as shown in the next figure.

\vskip.5cm
\centerline{\resizebox{10cm}{!}{\includegraphics{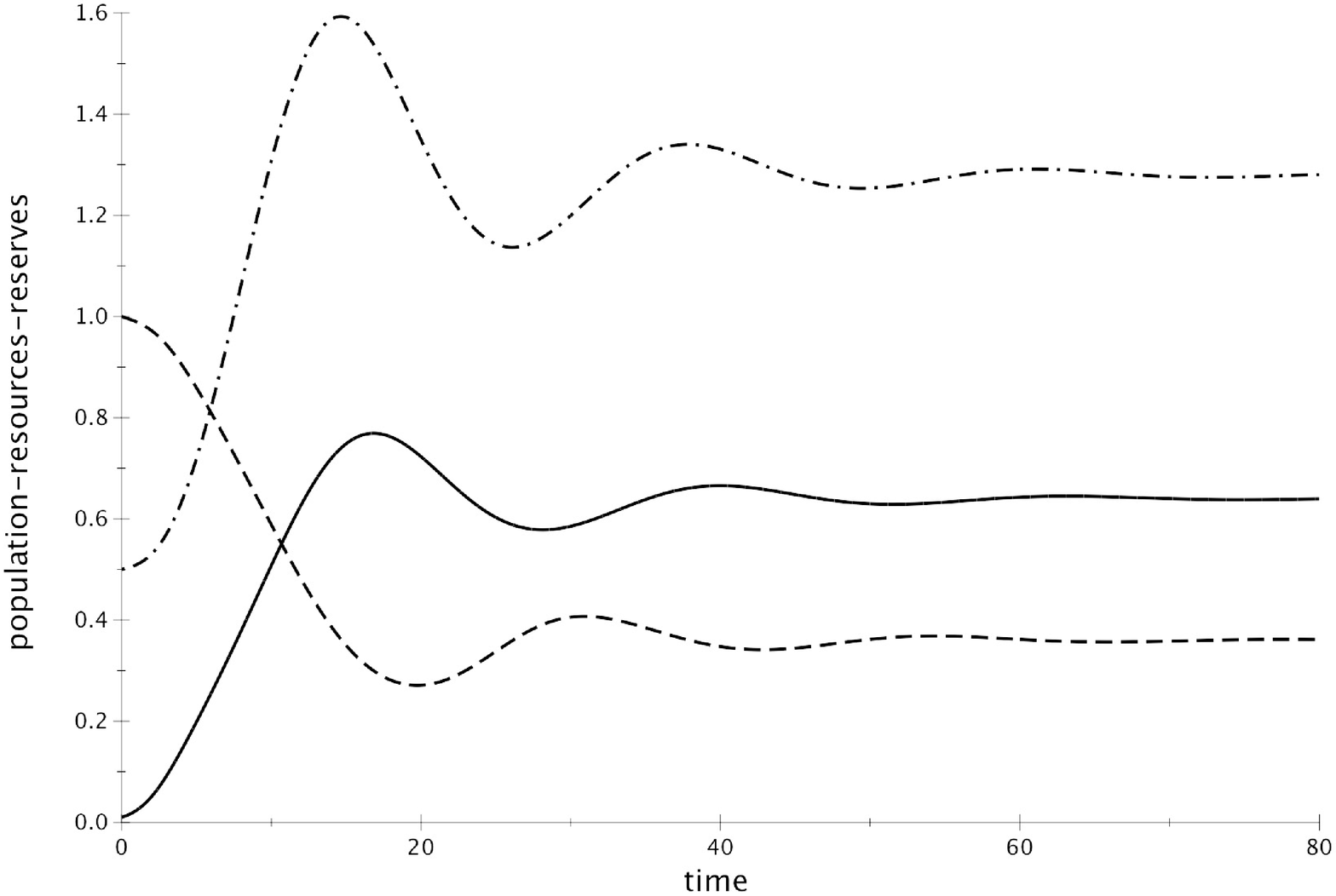}}}\vskip-.25cm\qquad\centerline{{Figure 7. Simulation results for system (\zoct) for $\alpha=1.0, \beta=2, \gamma=0.3, \delta=0.1, \lambda=2$ and $\mu=2$.}}\vskip.5cm

One interesting question one can ask at this point is whether it is possible, by choosing the parameters so as not to satisfy (\negcond), to have a solution displaying a limit cycle. It turns out that this is indeed possible as shown in Figure 8.

\vskip.5cm
\centerline{\resizebox{10cm}{!}{\includegraphics{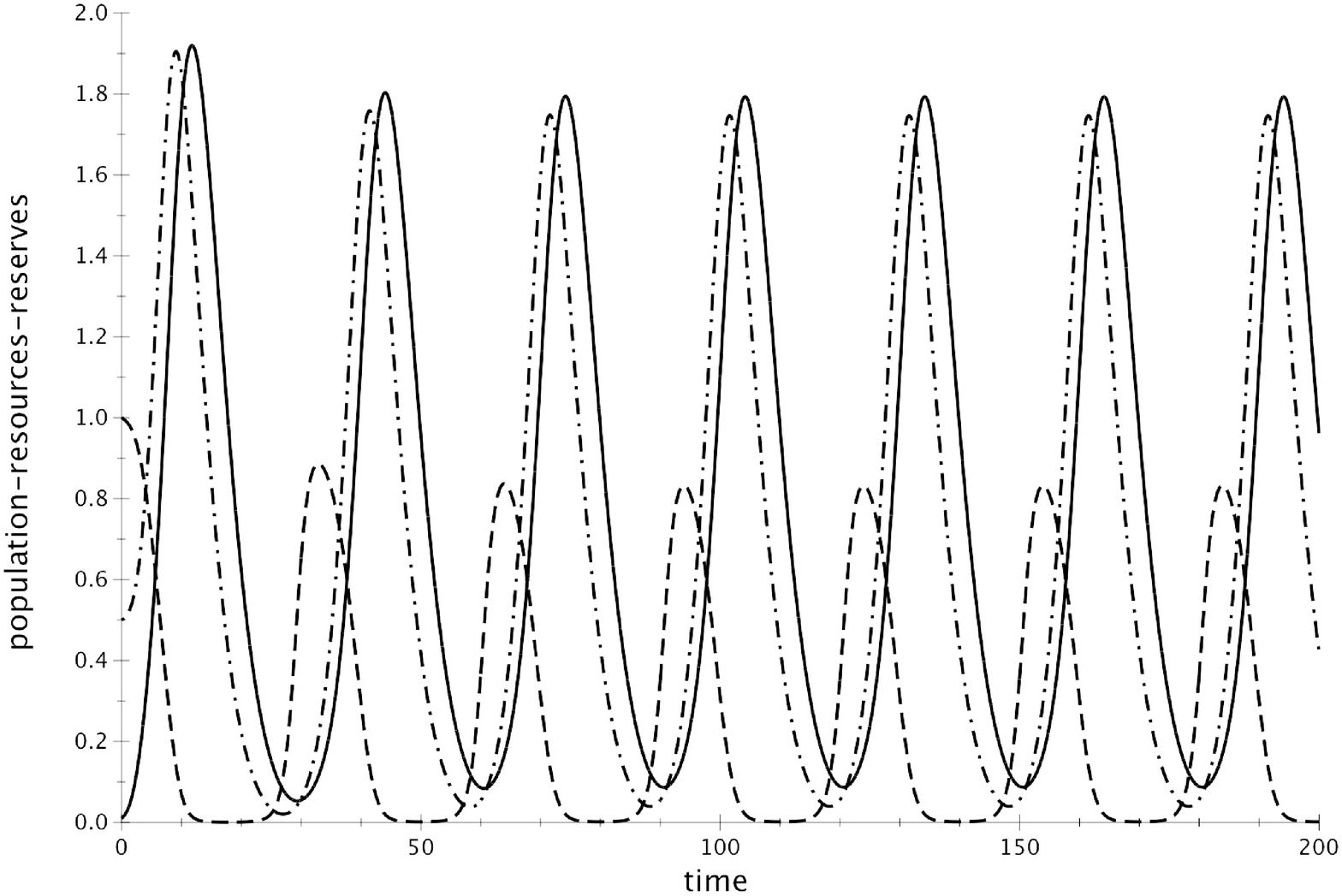}}}\vskip-.25cm\qquad\centerline{{Figure 8. Simulation results for system (\zoct) for $\alpha=1.25, \beta=1, \gamma=0.1, \delta=0.1, \lambda=2$ and $\mu=2$.}}\vskip.5cm

We do not wish to assign any metaphysical meaning --in the vein of ``history repeating itself''-- to this result, but merely include it here as a possibility that is present in our model. 

The final simulations were carried out using the exponential parametrisation for the functions $A$ and $B$ that give rise to system (\zenn). The same overall behaviour is present here, namely, depending on the parameters, one can have a collapse, reach a steady-state or enter a limit cycle.

A more interesting result however is shown in Figure 9. 

\vskip.5cm
\centerline{\resizebox{10cm}{!}{\includegraphics{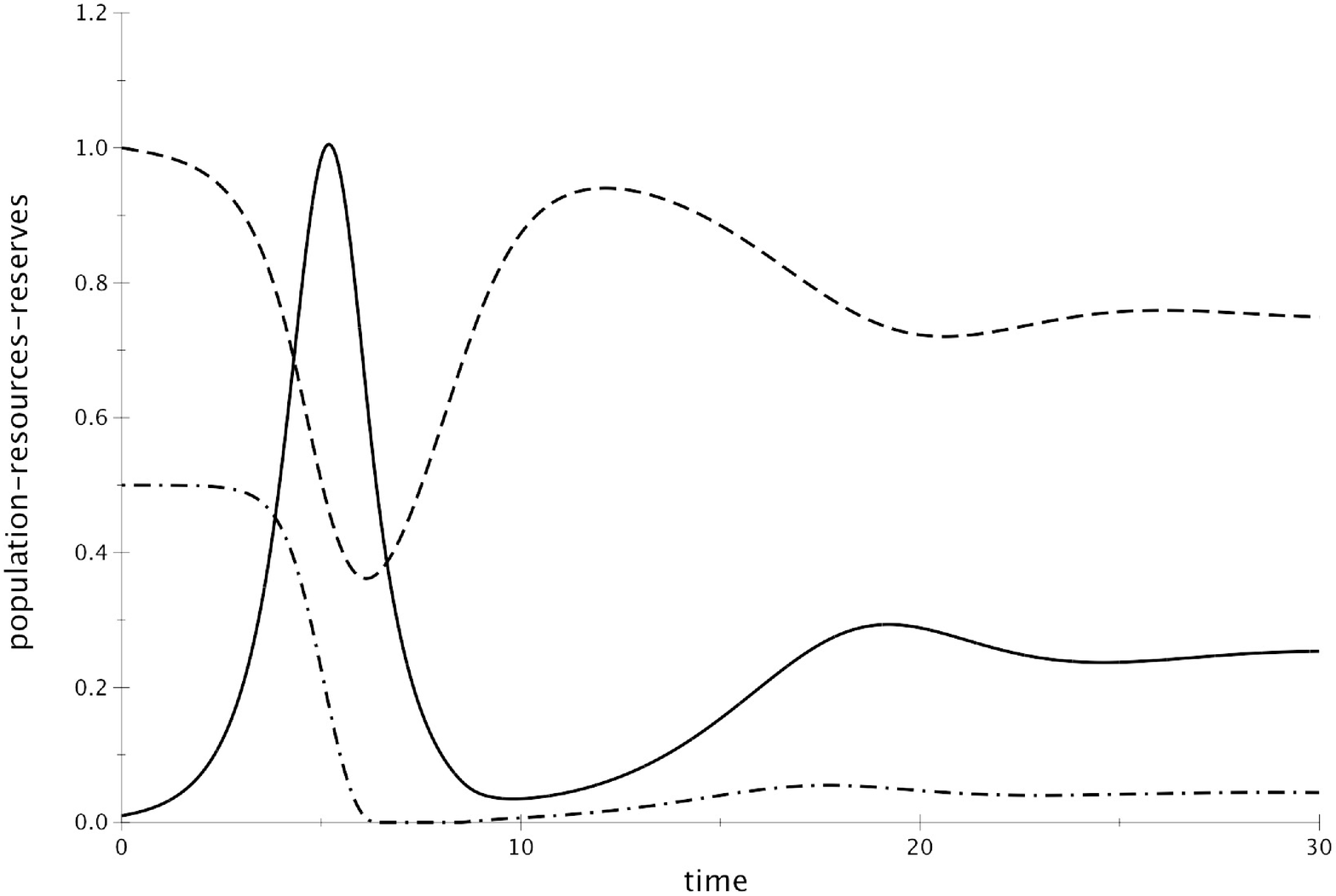}}}\vskip-.25cm\qquad\centerline{{Figure 9. Simulation results for system (\zenn) for $\alpha=1, \beta=2, \gamma=1, \delta=0.5, \lambda=4$ and $\mu=4$.}}\vskip.5cm

Here we have a near-collapse situation, where all non-renewable reserves are exhausted. Still the population, albeit hovering on the brink of extinction for some time, manages to pick-up and grow again, reaching a steady-state eventually. In this case the new beginning is totally different from the initial one, since there are no pre-existing reserves and consumption can only use whatever is produced in a labour-intensive way. 

What is interesting in the present case, is that one can fine-tune the stiffness of the functions $A$ and $B$ and thus study what is known as the `Seneca cliff' or `Seneca effect' [\refdef\bardi] . This is a reference to the Roman stoic philosopher's  writings  in which he pointed out that ``increases are of sluggish growth, but the way to ruin is rapid". While initially conjectured, by  M.K. Hubbert [\refdef\hubbert], that a collapse follows a bell-shaped curve, the works of Forrester pointed towards an asymmetrical situation. It is therefore interesting to investigate the details of the collapse curve in the framework of our model. Figure 10 shows a typical collapse situation. 

\vskip.5cm
\centerline{\resizebox{10cm}{!}{\includegraphics{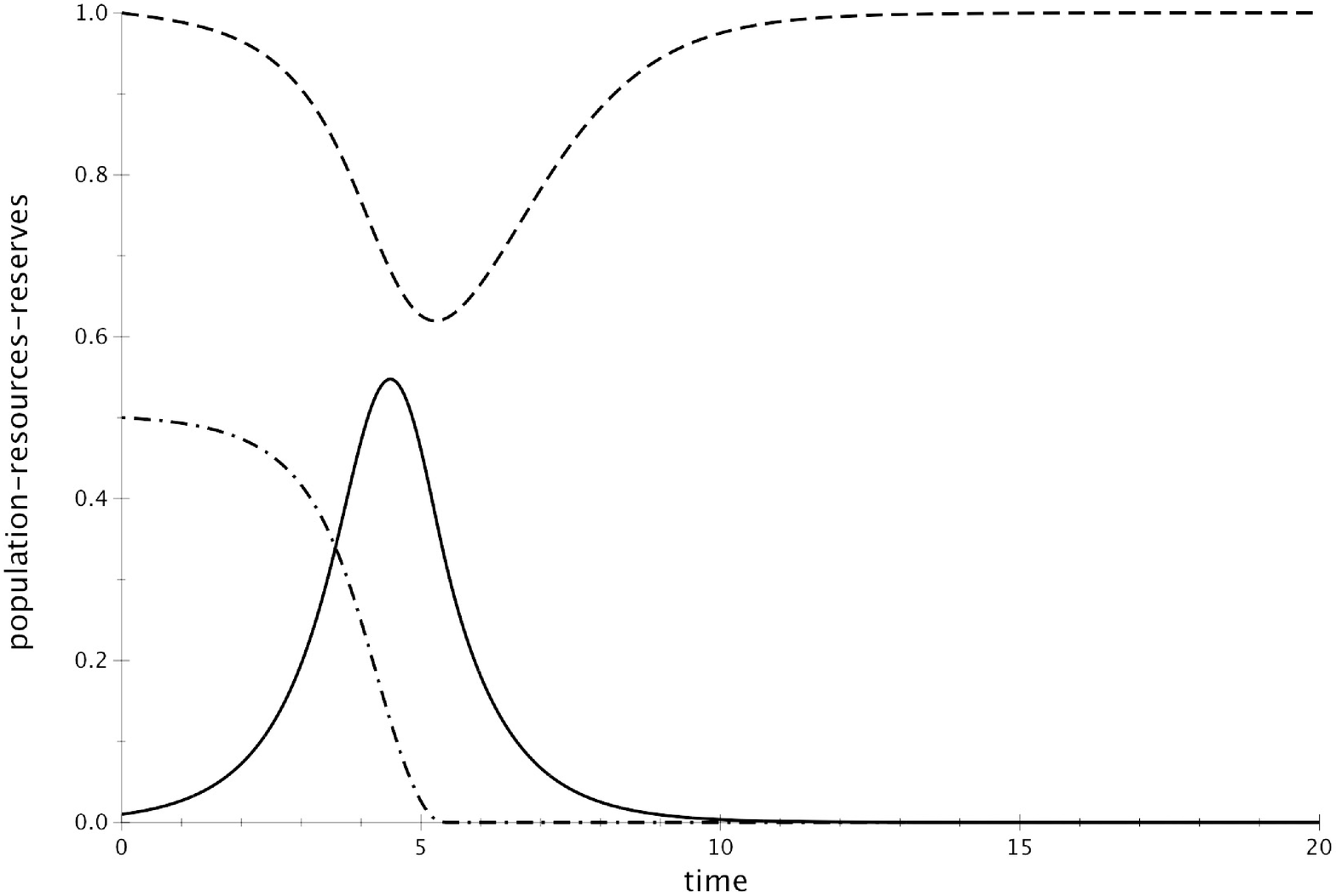}}}\vskip-.25cm\qquad\centerline{{Figure 10. Simulation results for system (\zenn) for $\alpha=1, \beta=2, \gamma=1.4, \delta=0.5, \lambda=3$ and $\mu=3$.}}\vskip.5cm

Here the parameters were chosen so as to have a shape that is as symmetrical as possible. Increasing the value of $\beta$ in the parametrisation of the function $A$ (and tweaking slightly the values of $\lambda$ et $\mu$) it is possible to obtain a strongly asymmetrical collapse in the form of a Seneca cliff.  Figure 11 shows just such a situation. 

\vskip.5cm
\centerline{\resizebox{10cm}{!}{\includegraphics{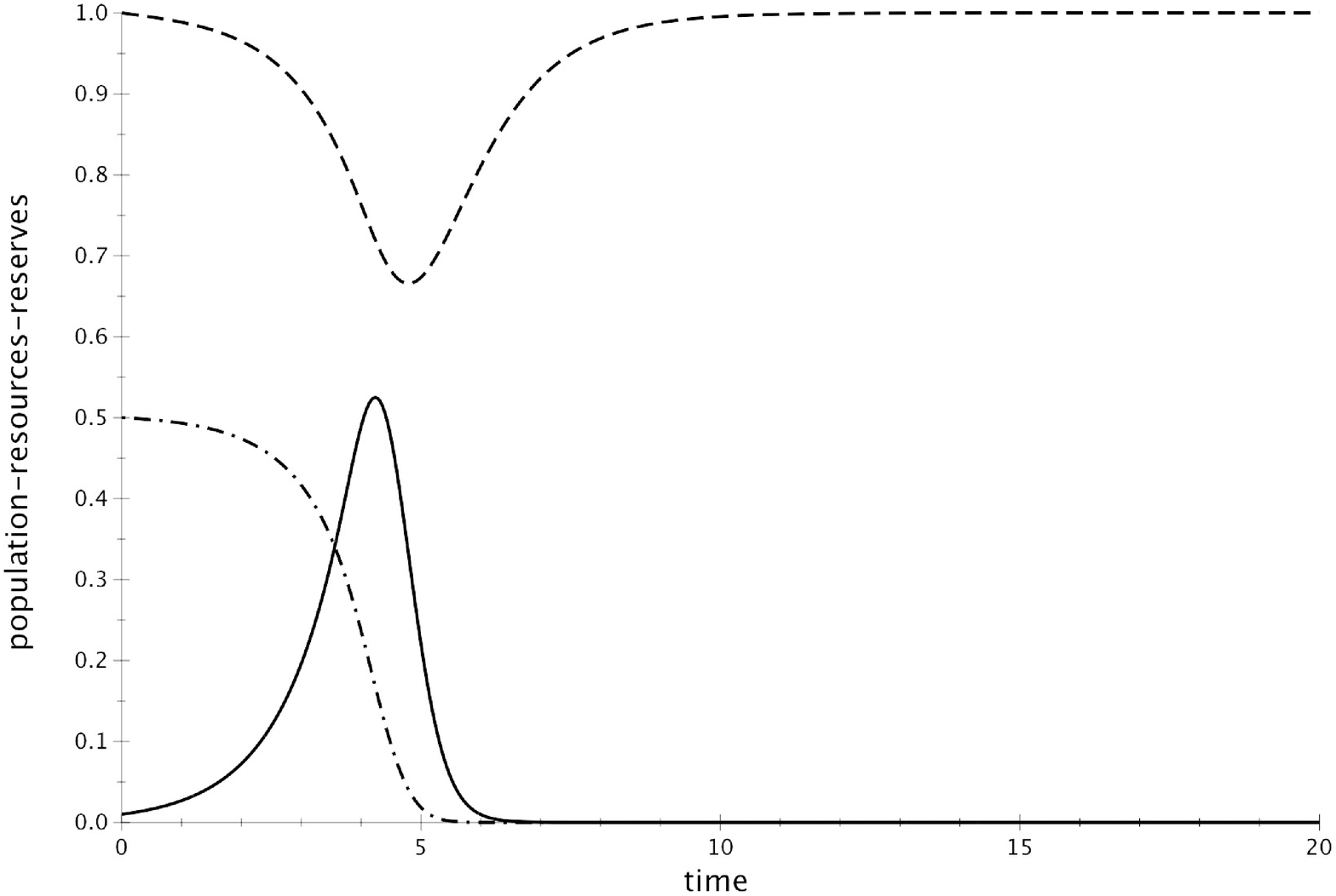}}}\vskip-.25cm\qquad\centerline{{Figure 11. Simulation results for system (\zenn) for $\alpha=1, \beta=5, \gamma=1.4, \delta=0.5, \lambda=5$ and $\mu=5$.}}\vskip.5cm

Thus not only the existence of a steady-state vs. the appearance of a collapse is conditioned by the choice of parameters of the model, but also the details of the collapse itself, to say nothing of the route to a steady state. Clearly this situation raises several questions which we shall address  in the conclusions of this article. Note also that it appears to be impossible to obtain collapses such as those in Figures 10 and 11 under the positivity condition [C1], which explains why we insist on condition [C2] instead.

\bigskip
5. {\scap Large discretisation steps and the ultradiscrete limit}
\medskip

In the introduction we mentioned that, in our view, for a discretisation scheme to be an appropriate one it should lead to sensible results even when the step is not very small [\stefan], i.e. even when we move away from the continuum limit. We claim that all the discrete schemes proposed in this paper do satisfy this requirement. The best way to illustrate this is by exhibiting the results of simulations performed with various choices of the step [\refdef\cycles]. The three figures that follow present such results for the model described by the discrete system (\zoct) obtained with discretisation steps 0.01, 1. and 100. respectively. (The choice of parameters of the model, as far as the positivity conditions are concerned, is not crucial. In the situation at hand we are not interested in the actual physical relevance of the model but only in seeing whether the aspect of the solution of (\zoct) is maintained when the step increases by several orders of magnitude).

Figure 12 corresponds to a standard simulation with a step of 0.01.

\vskip.5cm
\centerline{\resizebox{10cm}{!}{\includegraphics{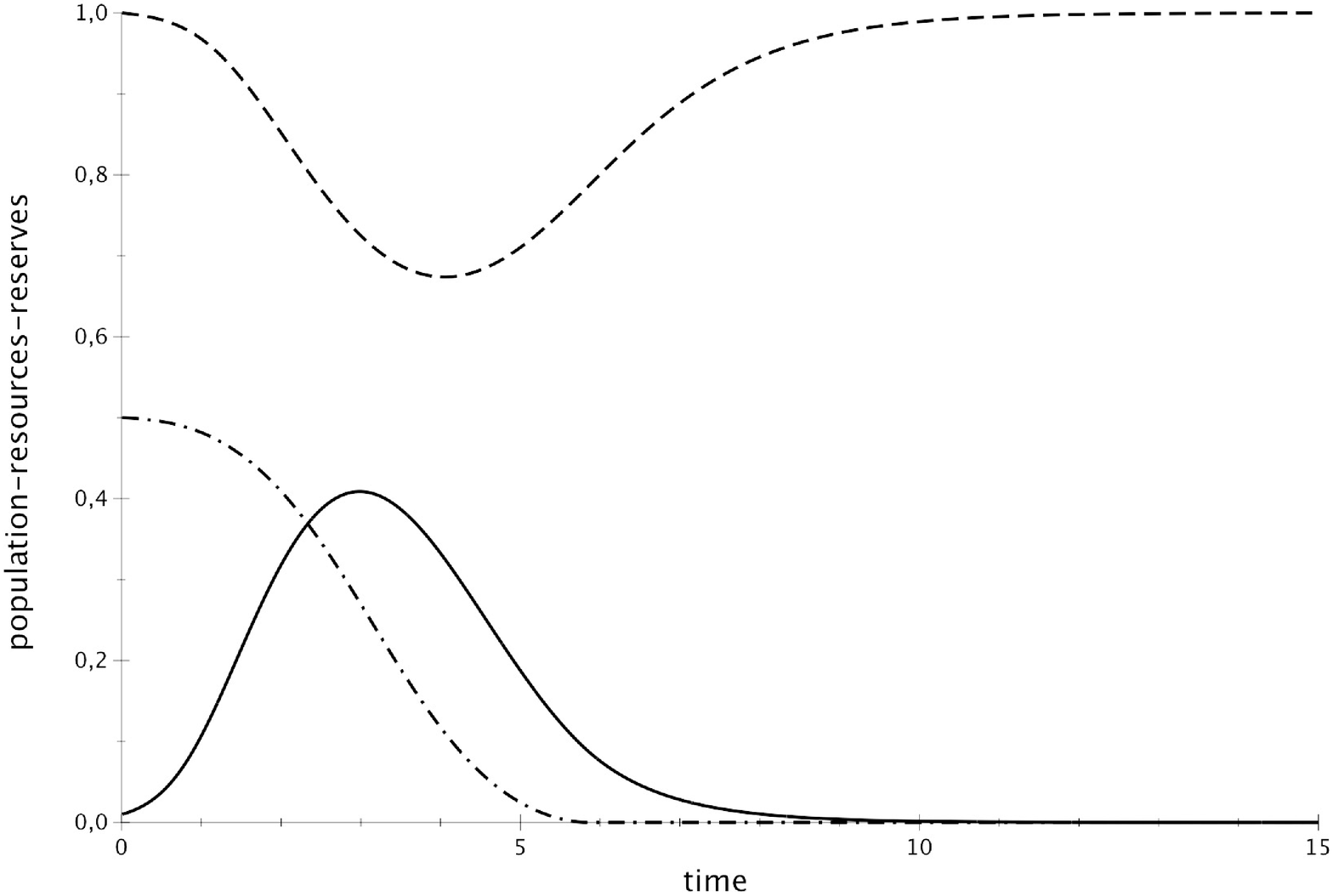}}}\vskip-.25cm\qquad\centerline{{Figure 12. Simulation results for system (\zoct)  for $\alpha=3.0, \beta=2, \gamma=1., \delta=0.5, \lambda=2$ and $\mu=2$ and $h=0.01$ .}}\vskip.5cm

Figure 13 represents the results of a simulation with the same parameters but with a step equal to 1.

\vskip.5cm
\centerline{\resizebox{10cm}{!}{\includegraphics{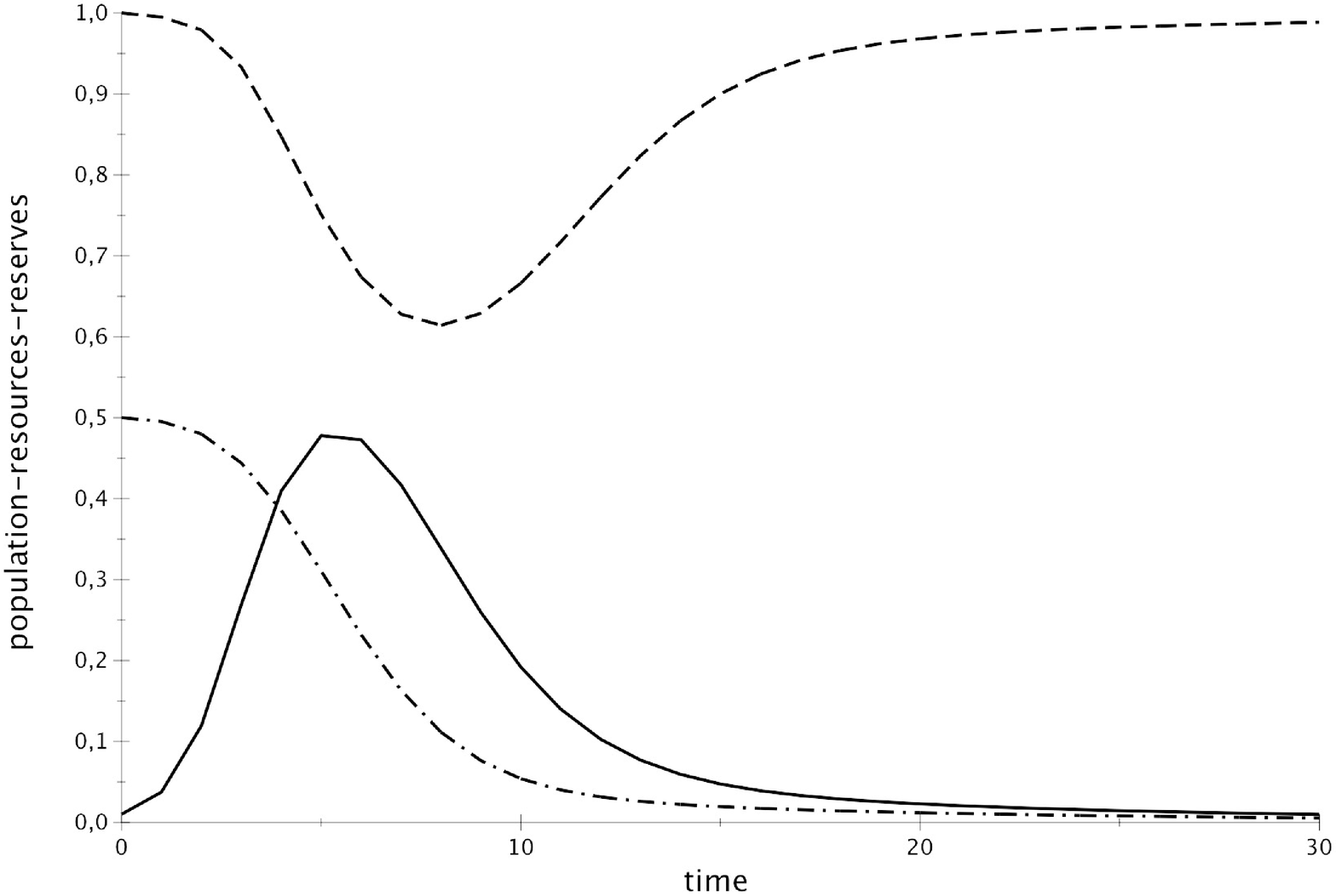}}}\vskip-.25cm\qquad\centerline{{Figure 13. Simulation results for system (\zoct)  for $\alpha=3.0, \beta=2, \gamma=1., \delta=0.5, \lambda=2$ and $\mu=2$ and $h=1$.}}\vskip.5cm

Finally Figure 14 is obtained for a step equal to 100.

\vskip.5cm
\centerline{\resizebox{10cm}{!}{\includegraphics{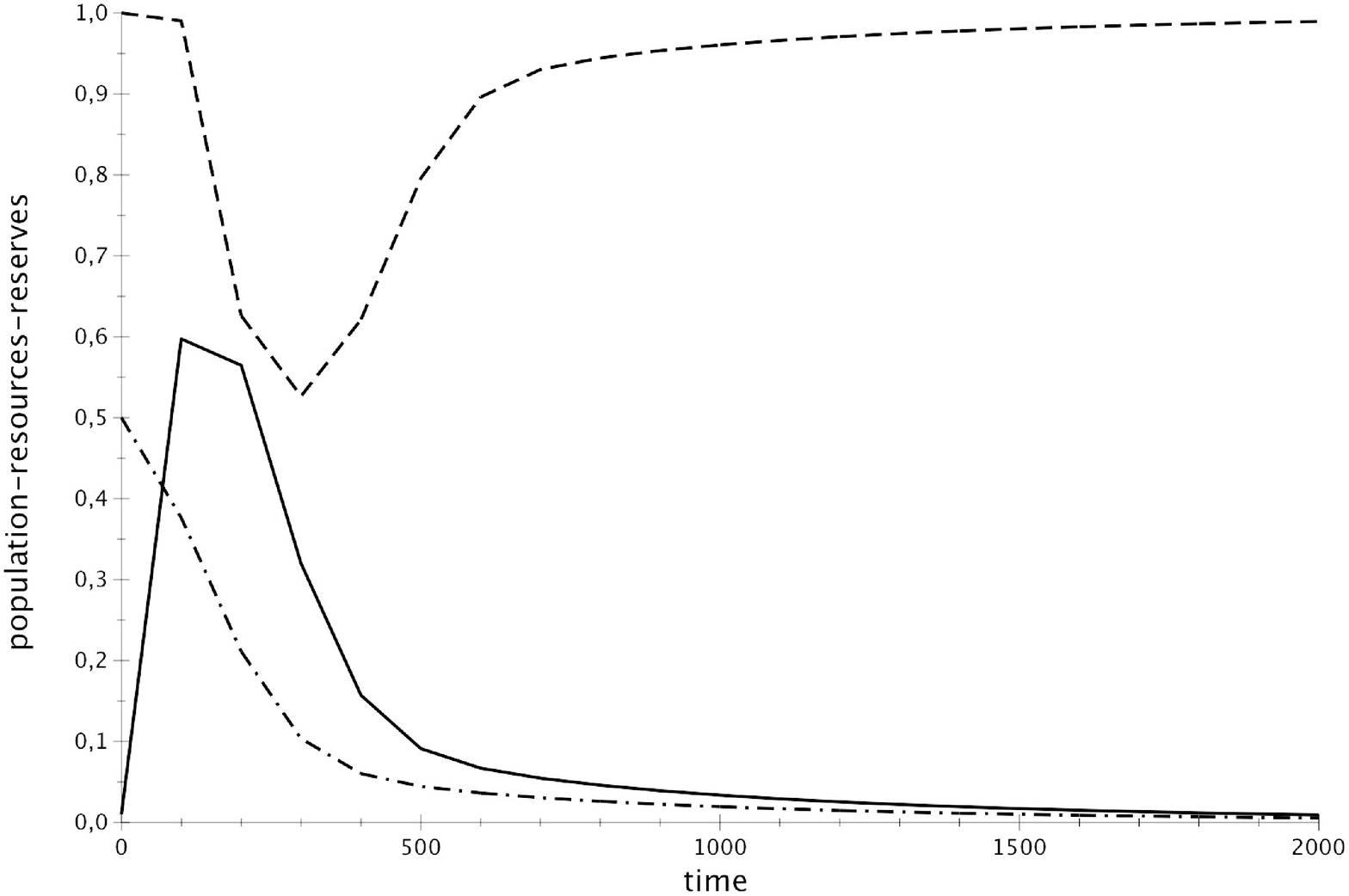}}}\vskip-.25cm\qquad\centerline{{Figure 14. Simulation results for system (\zoct)  for $\alpha=3.0, \beta=2, \gamma=1., \delta=0.5, \lambda=2$ and $\mu=2$ and $h=100$.}}\vskip.5cm

It is remarkable that the overall behaviour is maintained despite the immense increase of the discretisation step. And this remains true when one increase the step even further. So the question which arises naturally is whether going to the limit of an infinite step leads to dynamical equations which still make sense. As we shall show in what follows, this is indeed the case. The way to do this is by applying what is known as the ultra-discretisation procedure and obtain equations which are known [\ud] to define a generalised cellular automaton.

The interest in obtaining a cellular automaton through ultradicretisation is that it simplifies the dynamics, retaining only its essential features whereby making it, generally speaking, somewhat less sensitive to small changes in the parameters than the original continuous or discrete systems. Moreover, for such a cellular automaton, convergence to a fixed point at a finite location in phase space happens in a finite number of steps, which makes numerical simulations very fast. The application of the ultra-discretisation method is made possible thanks to the precautions we have taken when writing discrete equations that are devoid of any minus signs. Before proceeding further let us remind briefly how the procedure works.

One can think of the ultra-discrete (UD) limit as being the opposite of a continuum limit. The latter consists in taking the limit of the step $h\to0$ in the discrete equations. For the ultradiscrete limit we take $h\to\infty$. More precisely, we put $h=e^{1/d}-1$ ($d$ positive) and take the limit $d\to0$. Moreover we consider the logarithms of the dependent  variables, setting $hx=e^{X/d}$, $hy=e^{Y/d}$,  $h z=e^{Z/d}$, where $X, Z\in{\Bbb R}\cup \{-\infty\}$ and $Y\in\, \big]\!-\infty, 1\big] \cup \{-\infty\}$. For the parameters we set $h\alpha=e^{A/d}-1, h\beta=e^{B/d}-1$ 
etc. for $A,B,\dots>0$. The essential identity that allows us to write the equations at the limit $d\to0$ is $\lim_{d\to0}d\log(e^{A/d}+e^{B/d})=\max(A,B)$. Before presenting our detailed analysis of the ultra-discrete analogue of one of our models, it is interesting to present just the results of the simulation of the ultra-discrete system obtained from (\zoct), and compare them to those presented in Figures 12, 13 and 14.

\vskip.5cm
\centerline{\resizebox{10cm}{!}{\includegraphics{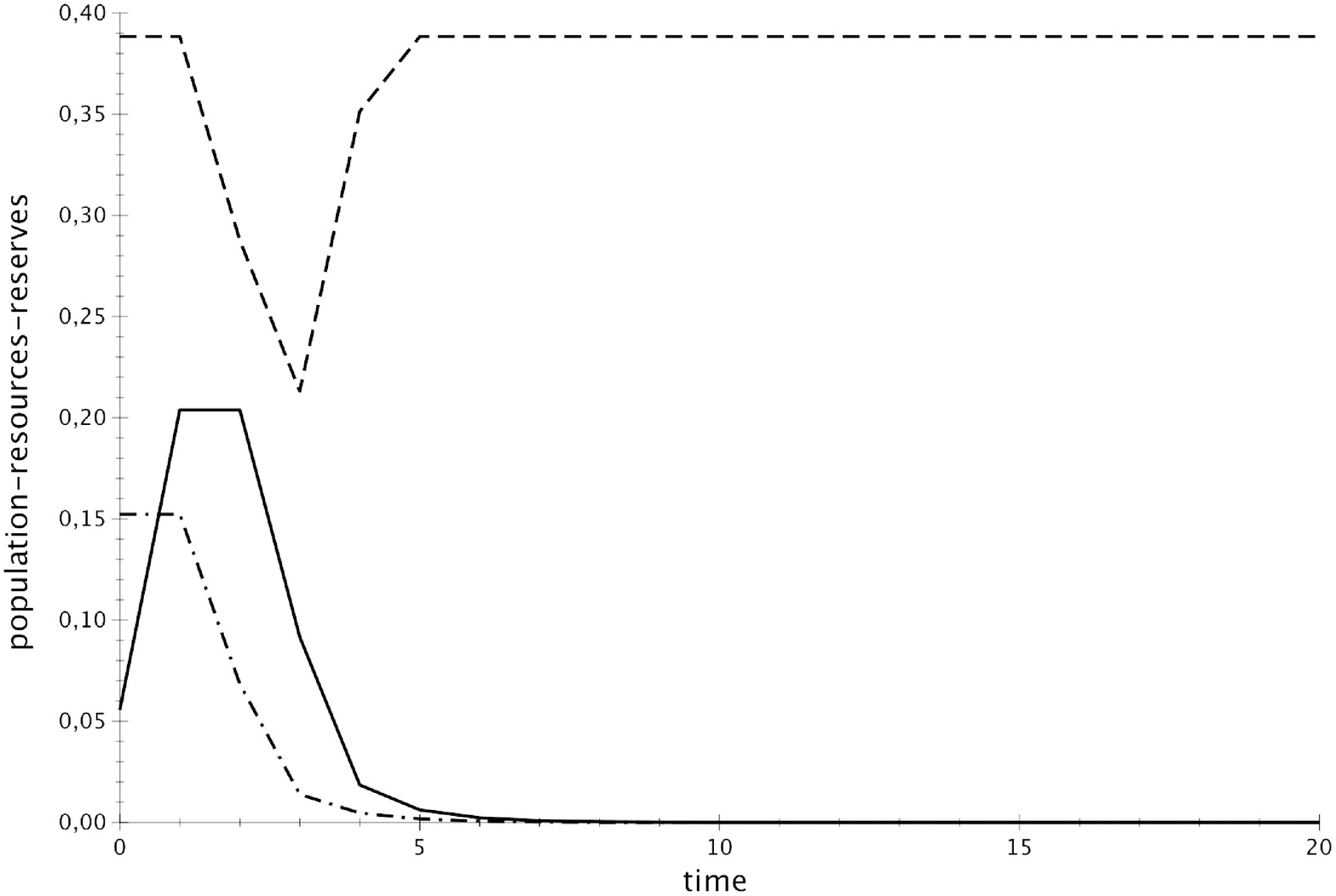}}}\vskip-.25cm\qquad\centerline{{Figure 15. Simulation results for the UD-limit of system  (\zoct) for $A=1.5, B=2.7, C=2.0 $ and $D=1.0$.}}\vskip.5cm

Figure 15 shows the evolution of  the exponential of the variables $X,Y,Z$ in arbitrary scale. Using the exponentials of the variables $X,Y,Z$ allows us to compare directly their behaviour to that of the variables $x,y,z$ of the discrete model. In fact, the similarity between figures 14 and 15 is striking.

Performing the detailed analysis of the ultra-discrete system associated with (\zoct) can be quite intricate. Thus, and for clarity's shake, we prefer to focus on the simplest among the models presented here, namely (\zhex).
Taking the ultradiscrete limit of the mapping (\zhex) we find thus 
$$X_{n+1}=\max(X_n,A+Z_n)-B\eqdaf\zenn$$
$$Y_{n+1}=Y_n+1-\max(0,X_n,Y_n)\eqno(\zenn b)$$
$$Z_{n+1}=Z_n+\max(Z_n,X_n+Y_{n+1})-\max(Z_n,C+X_n).\eqno(\zenn c)$$

The fixed point structure of (\zenn) is somewhat more complicated than that of (\zhex). In fact, we constructed the latter so as to have precisely the same fixed points as the continuous system (\ztes), but the structure of (\zenn) is dictated by the ultra-discretisation procedure. 
Let us first look at the conditions equation (\zenn a) imposes on a fixed point $(X_*, Y_*, Z_*)$. If $X_*$ takes a finite value, $B$ being positive, we find $Z_*=B-A+X_*$, implying that $Z_*$ necessarily takes a finite value as well. On the other hand, if $X_*=-\infty$ then necessarily $Z_*=-\infty$ as well.
From (\zenn b) we find that if $Y_*$ is finite then $X_*$ and $Y_*$ must satisfy $\max(X_*, Y_*)=1$. In particular, when $X_*=-\infty$ we have $Y_*=1$. If however $Y_*=-\infty$ then (\zenn b) does not impose any constraints on $X_*$. Similarly, in case $Z_*=-\infty$ we know that we must have $X_*=-\infty$ as well, and equation (\zenn c) does not furnish any information at all. However, when $Z_*$ takes a finite value (which, as we saw,  can only happen if $X_*$ is finite as well and in which case $Z_*=B-A+X_*$) we find from (\zenn c) that $\max(B-A, Y_*) = \max(B-A,C)$. Here there are two cases to consider: either $B-A\geq C$ or $B-A<C$. In the former case we find that $Y_*\leq B-A$ (which includes the possibility that $Y_*=-\infty$), whereas in the latter case we have that $Y_*=C$. Because of the conditions imposed on the variable $Y$, this last case is only possible if $C\leq 1$. 

In summary, we have the following fixed points for the ultra-discrete system (\zenn): $(-\infty, -\infty, -\infty)$ and $(-\infty, 1, -\infty)$ exist for arbitrary values of the parameters and, respectively, correspond to the fixed points $(0,0,0)$ and $(0,1,0)$ of the discrete (or continuous) system. In addition to these points there is the possibility of having a fixed point of the form $(X_*, Y_*, X_*+B-A)$ with $X_*, Y_*\in\, \big]\!-\infty, 1\big]$. When $B\geq A+C$ then $Y_*\leq B-A$ with $X_*$ and $Y_*$ such that $\max(X_*, Y_*)=1$. In particular, if $Y_*<1$ then $X_*=1$. However, when $Y_*=1$ we can only say that $X_*\leq1$ and, in general, it is impossible to predict the precise values of both $X_*$ and $Y_*$. When $B< A+ C$, for $C\leq 1$, we have that $Y_*=C$, which then fixes the value of $X_*$ to be 1 if $C<1$. On the other hand, when $C=1$, the precise value of $X_*$ is not determined a priori by the parameters of the ultra-discrete model.

It is clear that the fixed point $(X_*, Y_*, X_*+B-A)$ is an ultra-discrete counterpart of the fixed point $(1-\gamma, \gamma, \beta(1-\gamma)/\alpha)$ that we found for the discrete and continuous systems. This statement, however, requires some further elucidations.  First of all, notice that said fixed point includes the case $(1,-\infty, 1+ B-A)$, which would correspond to a fixed point for the discrete case with $y_*=0$ but $x_*\neq1$, which in reality cannot exist since we impose $\gamma>0$. However, as the ultradiscretisation procedure we use is a limiting concept, sharp inequalities vanish at the limit and one has to admit the possibility of having $Y_*=-\infty$.
A second important point concerns the conditions $B\geq A+C$ and $B<A+C$. For the continuous and discrete toy models (\ztes) and (\zhex) we imposed the conditions $\beta^2\geq 4 \alpha \gamma$ or $\beta^2\geq 4 \alpha (\gamma-1)$ which, due to the positivity of the parameters $A,B$ and $C$, both yield the constraint $2B\geq A+C$ at the ultra-discrete limit (except for the case of the latter condition with $\gamma\leq 1$ for which there is no constraint on $B$). 
On the other hand, for the discrete model (\zhex) we also imposed a constraint on the time step $h$, $\beta>\alpha\gamma h$, for the model to be a faithful integrator of the original continuous one. At the ultra-discrete limit this condition becomes $B\geq A+C$, which in fact implies the former constraint but not the other way around. 
The situation in the ultra-discrete case is of course rather different from the discrete one because we not only lost the discrete time step $h$ as an adjustable parameter, but we actually let $h$ tend to infinity whereas the role of the condition $\beta>\alpha\gamma h$ in the discrete model was to give an upper bound on the size of $h$.

From the above analysis we saw that when $B\geq A+C$ there can exist a fixed point of the above shape for which $Y_*\leq B-A$, but that the precise value of $Y_*$ cannot be determined. In fact, the model by itself does not guarantee that $Y_*\leq1$, i.e. that the fixed point actually exists. Moreover, for the discrete model we know that the fixed point $(1-\gamma, \gamma, \beta(1-\gamma)/\alpha)$ can only exist when $\gamma<1$, a constraint which at the ultra-discrete limit becomes $C\leq 1$. Hence, if for the ultra-discrete model (\zenn) we impose $1\geq B-A \geq C$ as a condition on the parameters, then the existence of the above fixed point {\sl is} guaranteed and, moreover, we can say that the ultra-discrete model indeed describes the behaviour of the discrete model we use as an integrator in the simulations in section 4.

On the other hand, when $B-A<C\leq1$, the fixed point still exists (just as it does in the discrete case even if $\beta\leq \alpha\gamma h)$, 
but the ultra-discrete system will now give a (simplified) description of the behaviour of the discrete model in what one could call the `discrete regime', i.e. for the discrete model {\sl per se}, and not only for the integrator of the continuous one. We believe that both regimes are worth investigating as they will give a full picture of the dynamics that can be expected in the discrete system (\zenn).

Since a local, linear, stability analysis is meaningless on the ultra-discrete system, we shall carry out  our analysis of (\zenn) by direct simulation. Again, in order to make it easier to visually assess the results we plot the exponentials of the variables $X,Y,Z$, which are expected to behave like the variables $x,y,z$ of the discrete model. 

Figure 16 shows a situation where all three quantities tend to a finite fixed point. Note that the asymptotic value is reached here after just two steps.  This makes it extremely easy to numerically verify the attractive nature of this equilibrium for a large set of initial conditions, thereby effectively verifying its global stability. Moreover, the initial conditions for the UD system being in a sense an avatar for generic, asymptotically far, initial conditions for the discrete system, one is tempted to surmise that the corresponding fixed point, for the discrete system, will be globally stable as well.

\vskip.5cm
\centerline{\resizebox{10cm}{!}{\includegraphics{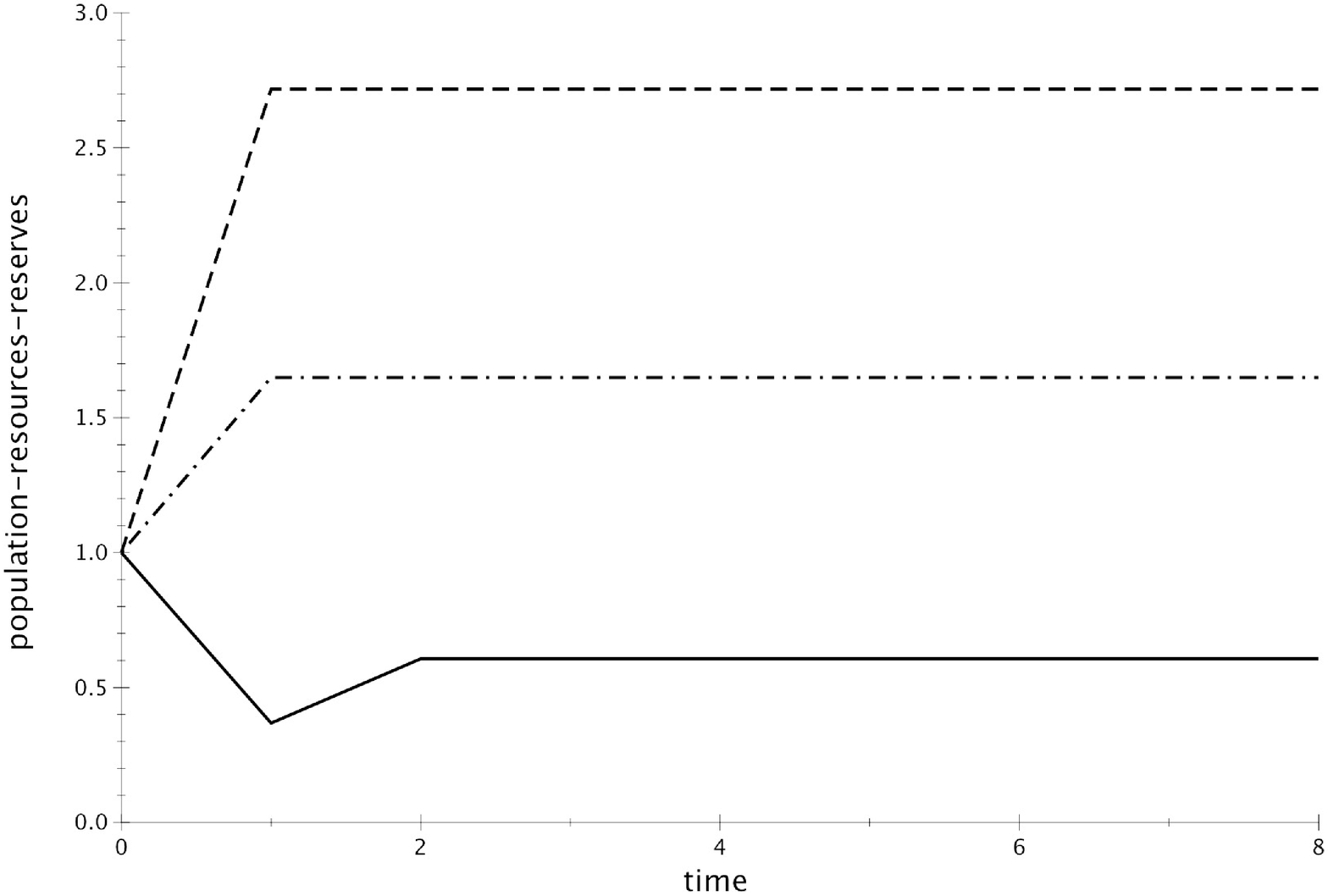}}}\vskip-.25cm\qquad\centerline{{Figure 16. Simulation results for the 
system (\zenn) with $A=1.5, B=2.5, C=0.5 $.}}\vskip.5cm

Figure 15, which we gave above,  corresponds to a collapse situation for the ultra-discrete analogue of (\zoct). Obviously similar situations exist in the case of system (\zenn). Figure 17 corresponds to such a case.

\vskip.5cm
\centerline{\resizebox{10cm}{!}{\includegraphics{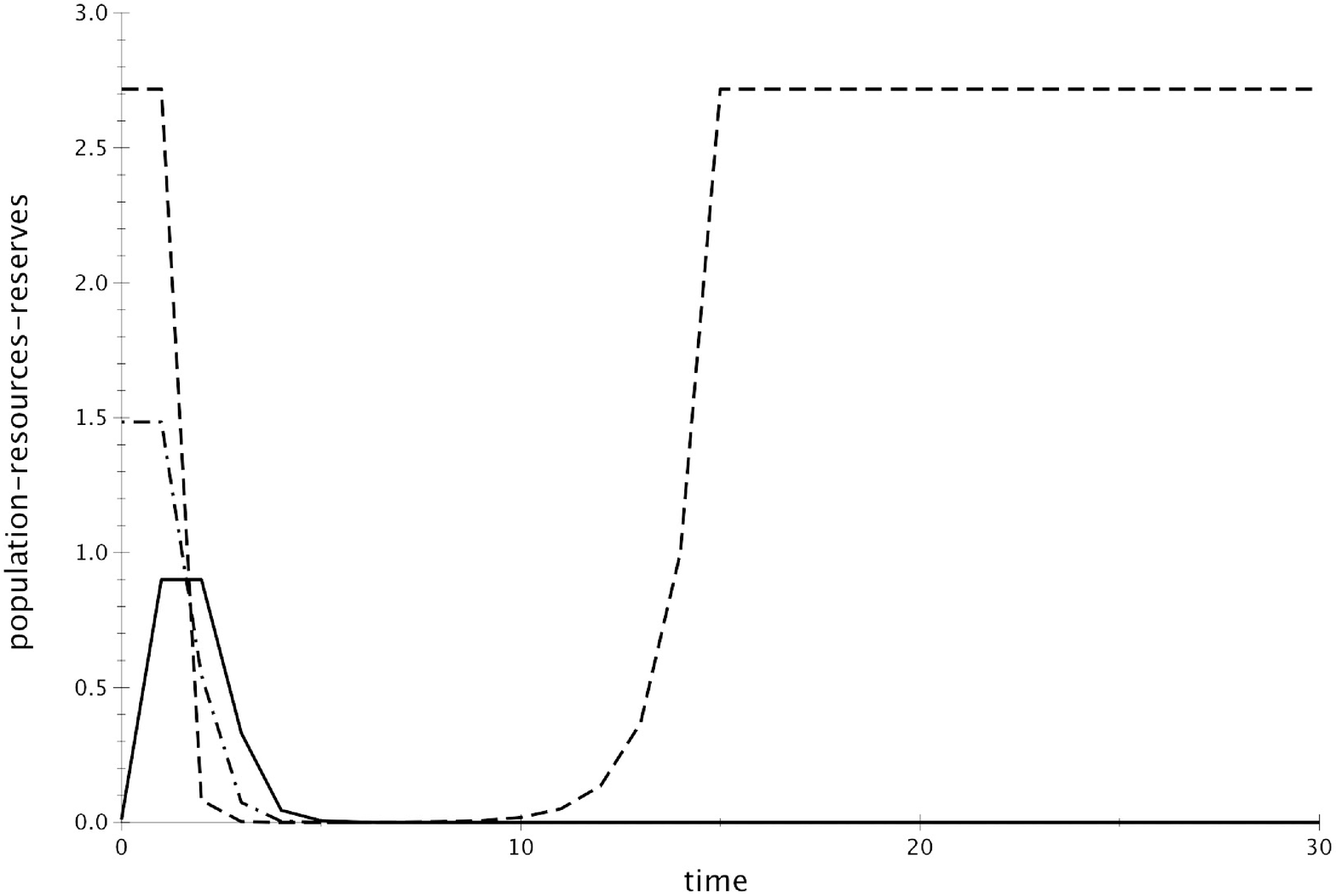}}}\vskip-.25cm\qquad\centerline{{Figure 17. Simulation results for the 
system (\zenn) with $A=1.5, B=2.0, C=1.5$.}}\vskip.5cm

But even more interesting is the case presented in Figure 18

\vskip.5cm
\centerline{\resizebox{10cm}{!}{\includegraphics{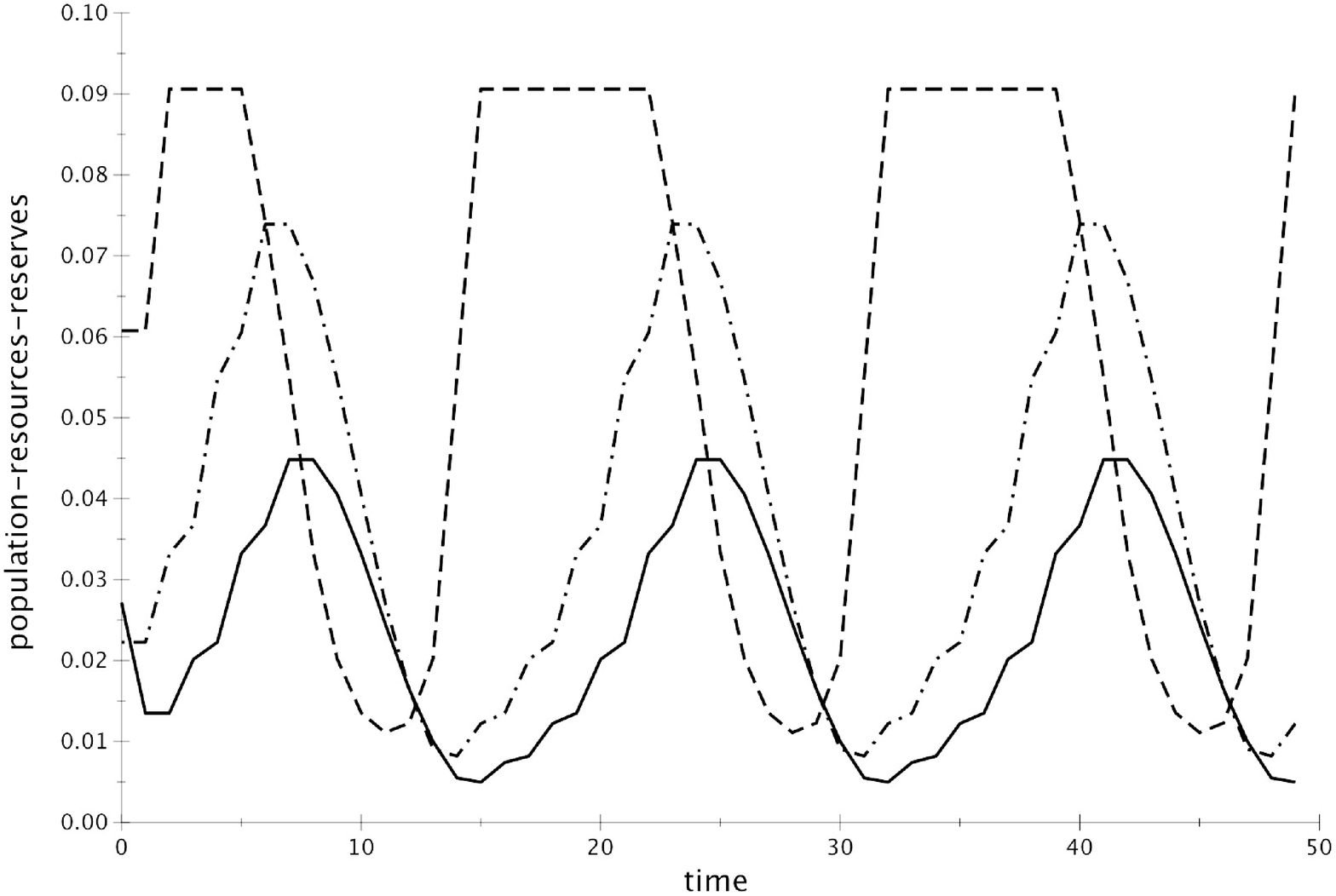}}}\vskip-.25cm\qquad\centerline{{Figure 18. Simulation results for the 
system (\zenn) with $A=1.5, B=2.0, C=0.6$.}}\vskip.5cm

Here the choice of the parameters is such that $C>B-A$ with $C<1$. Thus we can have $Y_{\star}=C$ and $X_{\star}=1$ as fixed point. However it turns out that in this case the system enters a limit cycle where, after a transient, all the variables oscillate periodically.

 It is interesting to notice that the cellular automaton leads to very rich dynamics. Contrary to the common belief that the cellular automaton divests the dynamics of most of its features (which are then, {\sl post facto}, considered to be non-essential) our results above show that most features of the dynamics are in fact conserved in the discrete to cellular automaton transition. This also goes to show that these dynamical structures, in the discrete system, are remarkably robust over vast parameter ranges.

\bigskip
6. {\scap Conclusions}
\medskip

In this paper we have presented a model inspired by the work of Motesharrei and collaborators on the HANDY model they proposed in [\handy] . Our aim was to introduce a genuinely minimal model while still retaining a rich dynamical behaviour. We started by constructing a differential model which we proceeded to discretise, following methods we have been developing over the years. In this way the discrete system could be considered as an integrator for the differential one and it was used in fact in all our simulations (thereby eliminating the need for a black-box numerical integrator). The fact that our discretisation method preserves positivity made possible the derivation of a cellular automaton analogue of our system, through the method of ultradiscretisation. 

Based on our study of the model, presented in the previous sections, the conclusions we can draw  are at two different levels. First, viewed purely as a dynamical system, despite its simplicity the model presented here turned out to be particularly rich in its dynamical behaviour. Depending on the choice of parameters and on the specific choice of the functions $A$ and $B$ in the model, one can have a fast or slow collapse, a march towards a steady state either monotonically or through oscillations, or a near collapse followed by resurgence. A sustained oscillatory regime, in the form of a limit cycle is also possible. By choosing appropriately the parameters of the model、 one can also reproduce the so-called Seneca effect of slow growth followed by a rapid collapse. The main dynamical behaviour of the model is also present in its cellular automaton, ultra-discrete, limit. 

The second set of conclusions concerns the model as a plausible representation of the possible outcomes of the interaction between humans and nature. Here, our conclusions are mitigated. What our analysis has shown is that everything depends crucially on the choice of parameters and on the specific choices for the population growth and reserve consumption functions. Thus, unless one has very precise knowledge of the parameters involved, one cannot pretend at realistic predictions. This is already true for our model which has an extremely restricted set of parameters. We do not believe that more sophisticated models involving dozens of variables, a multitude of coupling assumptions and hundreds of parameters can pretend to make surefire predictions about the future of the human species. What these models --but already a simple model like the one studied here as well-- do teach us is that a collapse of human civilisation is not impossible and that even if it were to be followed by a new resurgence, such a collapse would still lead to a complete transfiguration of the human/nature interaction. Therein lies the usefulness of these models.

\bigskip
{\scap Acknowledgements}
\medskip

The authors would like to acknowledge most stimulating discussions  with A. Ramani.
R. Willox would like to acknowledge support from the Japan Society for the Promotion of Science (JSPS),  through the JSPS grant: KAKENHI grant number 18K03355. 

\bigskip
{\scap References}
\medskip

 {[\forrest]} J. Forrester, {\sl World Dynamics}, Wright-Allen Press, Cambridge, 1971.
 
 {[\meadows]} D. Meadows, J. Randers, D. Meadows and W. Behrens, {\sl  The Limits to growth: A report for the Club of Rome's Project on the Predicament of Mankind}, Universe Books, New York, 1972. 

{[\malthus]} T. Malthus, {\sl An essay on the principle of population}, J. Johnson, London, 1798.

 {[\hardin]} G. Hardin, Science 162 (1968) 1243.
 
 {[\turner]} G. Turner, Global Environmental Change 18 (2008) 397.

 {[\brander]} J. Brander and M. Scott Taylor, The American Economic Review 88 (1998) 119.
 
 {[\handy]} S. Motesharrei, J. Rivas and E. Kalnay, Ecological Economics 101 (2014) 90.
 
 {[\mickens]} B. Grammaticos, A. Ramani, J. Satsuma, R. Willox, J. Math. Phys. 53 (2012) 023506.

 {[\ud]} T. Tokihiro, D. Takahashi, J. Matsukidaira and J. Satsuma, Phys. Rev. Lett. 29 (1996) 3247.

 {[\stefan]} R. Willox, B. Grammaticos, A.S. Carstea and A. Ramani, Physica A 328 (2003) 13.

 {[\holling]} C. Holling, Canadian Entomologist 91 (1959) 385.
 
 {[\bardi]} U. Bardi, {\sl The Seneca Effect: Why Growth is Slow but Collapse is Rapid}, Springer 2017.
 
 {[\hubbert]} M. King Hubbert, 1962, {\sl Energy Resources}, National Academy of Sciences, Publication 1000-D, 1962.
 
 {[\cycles]} R. Willox, A. Ramani, J. Satsuma and B. Grammaticos, Physica A 385 (2007) 473.

\end{document}